\documentclass[%
reprint, 
 superscriptaddress,
 showkeys,
 amsmath,amssymb,
 aps, physrev,
]{revtex4-2}

\usepackage{xcolor}
\usepackage{upgreek}
\usepackage{mathtools}
\usepackage{xr-hyper}
\usepackage{seqsplit}

\usepackage{graphicx}
\usepackage{dcolumn}
\usepackage{bm}

\newcommand{\pa}[1]{{\color{orange}{#1}}}

\newcommand{\curlyL}{\mathcal{L}}

\begin{document}

\preprint{APS/123-QED}

\title[]{Spontaneous Emergence of Solitary Waves in Active Flow Networks with Elastic Elements}

\author{\underline{Rodrigo Fernández-Quevedo García}}
  \email{rodfer05@ucm.es}
 \affiliation{
 Departamento de Estructura de la Materia, Física Térmica y Electrónica, Universidad Complutense de Madrid, 28040 Madrid, Spain
 }
 \affiliation{
 Grupo Interdisciplinar de Sistemas Complejos, Universidad Complutense de Madrid, 28040 Madrid, Spain
 }%

\author{\underline{Gonçalo Cruz Antunes}}%
 \email{g.antunes@tu-berlin.de}
 \affiliation{
 Helmholtz-Institut Erlangen-Nürnberg für Erneuerbare Energien (IET–2), Forschungszentrum Jülich, Cauerstr. 1, 91058 Erlangen, Germany
 }%
 \affiliation{
 Institute of Physics and Astronomy, Theory Division, Technische Universität Berlin, Hardenbergstrasse 36, 10623, Berlin, Germany
 }
\author{Jens Harting}
  \affiliation{
  Helmholtz-Institut Erlangen-Nürnberg für Erneuerbare Energien (IET–2), Forschungszentrum Jülich, Cauerstr. 1, 91058 Erlangen, Germany
  }%
  \affiliation{
  Department Chemie- und Bioingenieurwesen und Department Physik, Friedrich-Alexander-Universität Erlangen-Nürnberg,
Fürther Straße 248, 90429 Nürnberg, Germany
}%

\author{Holger Stark}%
  \affiliation{
 Institute of Physics and Astronomy, Theory Division, Technische Universität Berlin, Hardenbergstrasse 36, 10623, Berlin, Germany
  }
  
\author{Chantal Valeriani}%
 \affiliation{
 Departamento de Estructura de la Materia, Física Térmica y Electrónica, Universidad Complutense de Madrid, 28040 Madrid, Spain
 }
 \affiliation{
 Grupo Interdisciplinar de Sistemas Complejos, Universidad Complutense de Madrid, 28040 Madrid, Spain
 }%

\author{Martin Brandenbourger}
\affiliation{
Aix Marseille Université, CNRS, Centrale Méditerranée, IRPHE, UMR 7342, Marseille, France
}%

\author{Juan José Mazo}
 \affiliation{
 Departamento de Estructura de la Materia, Física Térmica y Electrónica, Universidad Complutense de Madrid, 28040 Madrid, Spain
 }
 \affiliation{
 Grupo Interdisciplinar de Sistemas Complejos, Universidad Complutense de Madrid, 28040 Madrid, Spain
 }%

\author{Paolo Malgaretti}
  \affiliation{
  Helmholtz-Institut Erlangen-Nürnberg für Erneuerbare Energien (IET–2), Forschungszentrum Jülich, Cauerstr. 1, 91058 Erlangen, Germany
  }%

\author{Miguel Ruiz-García}
  \email{miguel.ruiz.garcia@ucm.es}
 \affiliation{
 Departamento de Estructura de la Materia, Física Térmica y Electrónica, Universidad Complutense de Madrid, 28040 Madrid, Spain
 }
 \affiliation{
 Grupo Interdisciplinar de Sistemas Complejos, Universidad Complutense de Madrid, 28040 Madrid, Spain
 }%

\date{\today}

\begin{abstract}
Flow networks are fundamental for understanding systems such as animal and plant vasculature or power distribution grids. These networks can encode, transmit, and transform information embodied in the spatial and temporal distribution of their flows. In this work, we focus on a minimal yet physically grounded system that allows us to isolate the fundamental mechanisms by which active flow networks generate and regulate emergent dynamics capable of supporting information transmission. The system is composed of active units that pump fluid and elastic units that store volume. From first principles, we derive a discrete model—an active flow network—that enables the simulation of large systems with many interacting units. Numerically, we show that the pressure field can develop solitary waves, resulting in the spontaneous creation and transmission of localized packets of information stored in the physical properties of the flow. We characterize how these solitary waves emerge from disordered initial conditions in a one-dimensional network, and how their size and propagation speed depend on key system parameters. Finally, when the elastic units are coupled to their neighbors, the solitary waves exhibit even richer dynamics, with diverse shapes and finite lifetimes that display power-law behaviors that we can predict analytically. 
Together, these results show how simple fluidic elements can collectively create, shape and transport information, laying the foundations for understanding---and ultimately engineering---information processing in active flow systems.
\end{abstract}

\maketitle




Fluidic networks represent a unifying motif in the architecture of natural systems, providing efficient transport and dynamic regulation of matter and signaling across scales—from plant and animal vasculature to complex flow networks in mycorrhizal fungi~\cite{katifori2018transport, secomb2017blood, kramer2020pare, kramer2023biological, huizinga2009gut, moore2018lymphatic, alim2013random, alim2017mechanism, oyarte2025travelling}. In these naturally evolved networks, mechanisms such as periodic peristaltic waves in Physarum polycephalum~\cite{alim2013random, alim2017mechanism} and in the gut~\cite{huizinga2009gut}, or the coordinated contraction of elastic channels in lymphatic systems~\cite{moore2018lymphatic}, exemplify how soft, compliant fluid structures achieve robustness, adaptability, and directional control. These principles have inspired artificial microfluidic architectures that now harness local or distributed actuation, including micropumps, soft valves, and nonlinear hydraulic resistances~\cite{stone2009tuned-in,mosadegh2010integrated,duncan2013pneumatic,rothemund2018soft,case2019braess,preston2019soft,preston2019digital,brandenbourger2020tunable,ruiz2021emergent,ruiz2020topologically,martinez2024fluidic,jones2021bubble,decker2022programmable,li2022soft,kong2017open,metafluidics_web_1,van2023nonlinear,baeyens2023fons, jones2023soft, mosadegh2014pneumatic,thampi2016active, leyva2025active, winn2026unidirectionalflowcontinuousbroken}.

In this work, we study active flow networks—systems of interconnected channels in which flows are driven locally, rather than imposed at the boundaries by an external pump. Biological systems such as Physarum polycephalum~\cite{alim2013random, alim2017mechanism} or mycorrhizal fungi~\cite{oyarte2025travelling} illustrate the ubiquity of active flow networks in nature. Recently, interest in this field has grown significantly, motivating both the experimental realization of synthetic active flow networks in the laboratory \cite{jorge2024active, jorge2025active} and the development of theoretical models to describe their dynamics~\cite{ Woodhouse2016, souslov2017topological, Woodhouse2018, forrow2017mode}.

A defining feature of these fluidic systems, is their capacity to regulate signal processing and information transfer. By modulating flows~\cite{stone2009tuned-in,duncan2013pneumatic, case2019braess, brandenbourger2020tunable, rothemund2018soft,mosadegh2010integrated,leyva2025active, winn2026unidirectionalflowcontinuousbroken}, establishing pressure gradients~\cite{martinez2024fluidic, preston2019soft}, and directing localized transport events~\cite{ruiz2021emergent,ruiz2020topologically}, these networks can encode, transmit, and transform information embodied by the spatial and temporal distribution of chemicals~\cite{Walter2023}, pressure~\cite{duncan2013pneumatic,preston2019soft,decker2022programmable}, and velocity~\cite{thampi2016active,case2019braess}. At the local scale, this transduction is exemplified by active phoretic pumps~\cite{Antunes2022, Antunes2023, Michelin2019, Michelin2015,Yu2020,tan2019microfluidic}, which generate directed fluid motion through interfacial gradients---typically chemical, electrical, or thermal---resulting in surface-driven slip flows. Notably, such systems can exhibit hysteretic flow behavior~\cite{Antunes2023}, in which the directionality of transport depends on the history of pressure gradients, effectively encoding a memory of prior states. This raises a fundamental question: how do these two minimal ingredients---active flow and hysteretic memory---interact to regulate fluid transport in a network, and what new modes of information processing might they enable?

In this paper, we investigate a one-dimensional flow network comprising a series of active pumps interconnected by compliant channels, and examine the flow dynamics that emerge from the interplay of distributed hysteretic active units. Remarkably, this minimal fluidic circuit gives rise to active solitary waves, which are strikingly different from the behavior of isolated constitutive units, suggesting that such active networks may serve as a paradigmatic example of fluidic metamaterials~\cite{zhang2018metafluidic,kong2017open,metafluidics_web_1,metafluidics_web_2, djellouli2024shell, sehgal2024programmable,ruiz2021emergent,ruiz2020topologically,martinez2024fluidic}. These waves are robust, self-sustaining, and constitute discrete information packets in the form of traveling volume accumulations coupled with complex flow patterns. Remarkably, this represents a novel mode of transport in which---unlike previous approaches that rely on complex external driving---the speed and trajectory of localized information packets are regulated intrinsically by local dynamics.

\section{Networks of catalytically active channels interspersed with elastic units}

\begin{figure*}
\centering
\includegraphics[width=\linewidth]{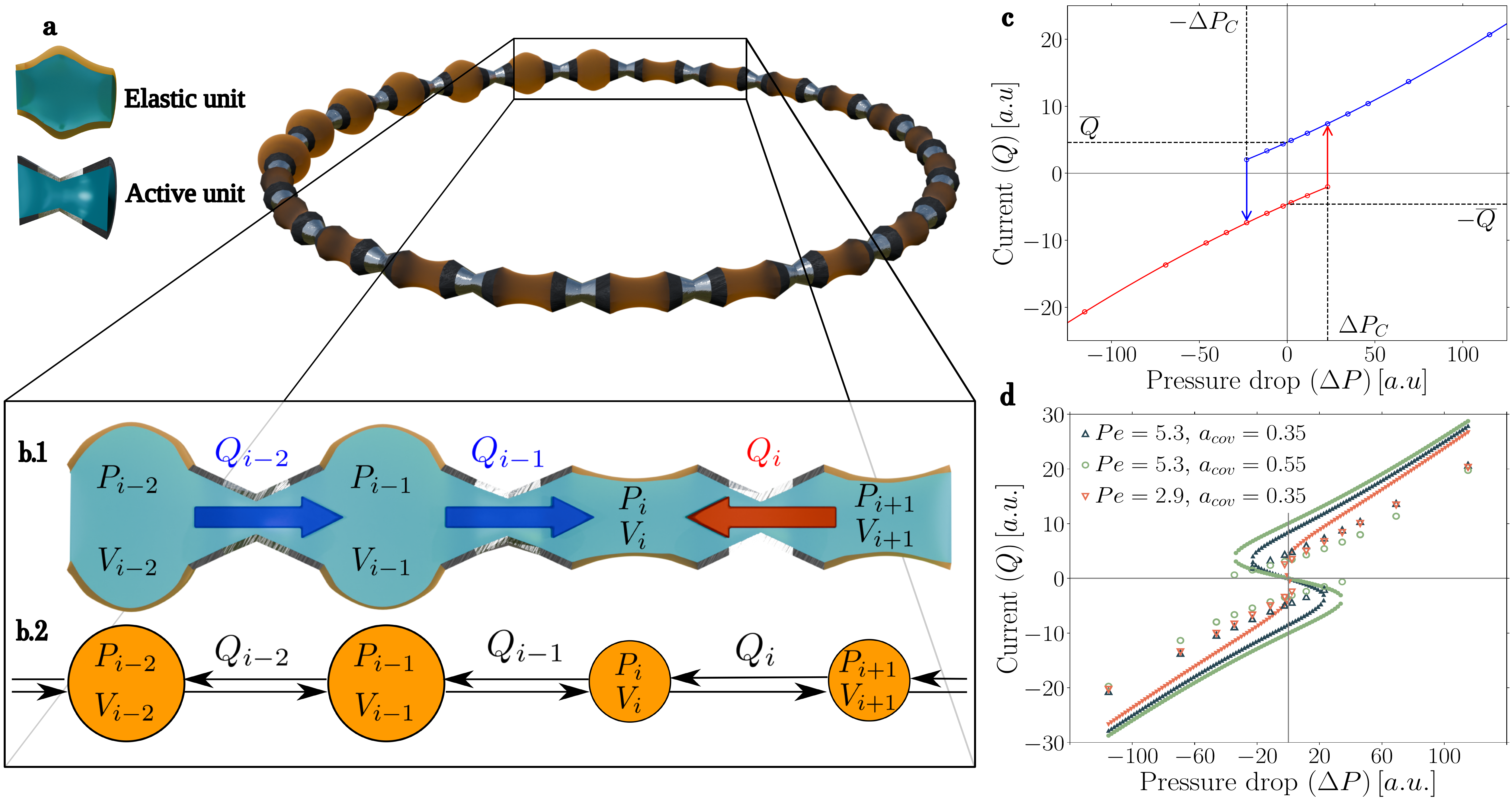}
\caption{Schematic representation of the active flow network. Panel \textbf{a} shows a ring that is composed of active units (regions bounded by black and gray walls) and elastic units (regions enclosed by brown walls), with some elastic units being larger than others, indicating the presence of a propagating solitary wave. Panel \textbf{b.1} displays the cross-sectional view of the ring at the boundary of a solitary wave, with the fluid inside represented in light blue. Arrows denote the flow through the active units, going in and out the elastic units. The ring can also be represented as a flow network model as shown in panel \textbf{b.2}, where active units (arrows) and elastic units (orange circles) are labeled with the index $i$. Panel \textbf{c} presents the characteristic behavior of one active unit  (current versus pressure drop in arbitrary units). At active unit $i$ pressure drop is computed as $\Delta P_i =  P_i - P_{i+1}$. Open circles correspond to results from full hydrodynamic simulations (lattice Boltzmann method), while solid lines represent a quadratic fit according to Eq.~\ref{eq:currentDeltaP} (\textcolor{black}{$\overline{Q} \approx 4.61, m \approx 0.117$, $a \approx 2\cdot 10^{-4}$ and $\Delta P_c\approx23.04$}). Red and blue indicate the two possible branches in Eq.~\ref{eq:currentDeltaP}, namely positive (clockwise, $b_i=1$) in blue and negative (counterclockwise, $b_i=-1$) in red. Panel \textbf{d} presents full hydrodynamic simulations (open symbols) and semi-analytical theory (closed symbols) in arbitrary units, using two different P\'eclet numbers ($Pe$) for the solute, and two different extents of the catalytic coating ($a_{cov}$). Panel \textbf{c} corresponds to the case: $Pe=5.3$ and $a_{cov}=0.35$. Please see Supplementary Material \ref{sec:LBM} for the details of the full hydrodynamic simulations and semi-analytical theory, as well as the definition of the arbitrary units in the plots.}
\label{Fig_schematic_ring}
\end{figure*}

We study an active flow network composed of passive compliant elements interspersed by varying-section active phoretic pumps which catalyze a chemical reaction at their bottleneck (see Figs.~\ref{Fig_schematic_ring} \textbf{a} and \ref{Fig_schematic_ring} \textbf{b.1})~\footnote{In typical experimental implementations, the catalyst is platinum, the fuel is hydrogen peroxide, and the solute is oxygen \cite{Howse2007, Ebbens2010,Ebbens2012}}. Since the chemical reaction occurs only in the vicinity of the bottleneck, the concentrations of reactants and products are inhomogeneous hence leading to phoretic motion of the solution~\cite{Antunes2022, Antunes2023, Yu2020, Michelin2015, Michelin2019}. In particular, when the advection of reactants and products is strong enough, the local phoresis can lead to a spontaneous symmetry breaking and the onset of a net flow across the active element~\cite{Antunes2022, Antunes2023}.

The net flow set by the active phoretic pump upon varying an external pressure difference, $\Delta P$, is shown in Fig. \ref{Fig_schematic_ring} \textbf{c}. 
At $\Delta P = 0$, the active unit can pump fluid with equal probability in the positive or negative direction (blue or red arrows in Fig. \ref{Fig_schematic_ring} \textbf{b.1}), thus the flow presents a clear hysteresis when approaching $\Delta P = 0$ from large negative or positive pressures. This is a marked difference with respect to a passive channel, as the {current/}flow rate in the active unit is non-zero at zero pressure drop and it displays a hysteretic behavior as a function of pressure drop.
The elastic units are chambers whose walls are composed by elastic membranes (see Fig.~\ref{Fig_schematic_ring} \textbf{a} and \textbf{b.1}). As such, their volume may change depending on the pressure of the fluid within, in contrast with the rigid-walled active units. Elastic units are passive, with the membrane shape being determined by mechanical equilibrium between the pressure inside the chamber, the elastic forces from the deformation of the membrane, and the pressure outside the chamber. 

While the combination of the two units can lead to networks of any topology, we shall focus on the simplest one: a ring-shaped network composed of interspersed active and elastic units. As such, every elastic unit is connected to an active unit on either side and vice-versa, see Fig.~\ref{Fig_schematic_ring}\textbf{a}. 
While a full numerical study of this system is possible, its high numerical cost will prevent us from a comprehensive characterization of the system, in particular at longer time scales. Accordingly, 
we derive  
an effective flow network model by 
coarse-graining the continuous description of both the active and elastic units (section \ref{sec:methods}, with extra details found in Supplementary Material \ref{sec:continuummodel}). The main assumptions in this derivation are: flow velocities are weak, such that one may neglect the inertial term in the Navier-Stokes equation, as well as the viscous drag applied on the walls of the elastic unit; the latter deforms slowly enough such that the inertial term may be neglected in the momentum conservation equation for the elastic material;  both elastic and active units are channels of high aspect ratio.
We support the theory for the active units with full hydrodynamics (lattice Boltzmann) simulations in Supplementary Material \ref{sec:LBM}. 
These assumptions result in two major conclusions, which form the basis of our coarse-grained theory: the pressure inside an elastic unit is approximately homogeneous, and is linearly proportional to the volume of the unit; the {current} inside an active unit varies with the pressure drop between the ends of the unit as per Fig. \ref{Fig_schematic_ring} \textbf{c}.  As shown in Fig. \ref{Fig_schematic_ring} \textbf{d}, via lattice Boltzmann simulations and a semi-analytical approach based on the lubrication and Fick-Jacobs theories, we further show that the {current} at zero pressure drop ($\overline{Q}$) and the width of the bistable domain ($2 \Delta P_c$) are tunable by changing e.g. the length of the active unit, or the extent of the catalytic coating. A discussion of typical experimental parameter values can be found in section \ref{methods:experimental_values}. In the following section, we coarse-grain the full chemoelastohydrodynamic problem to a flow network model.

\section{Coarse-graining to a flow network model}

\label{sec:methods}
We now present a continuous description of the active and elastic units, and coarse-grain it into an effective flow network model. Such a procedure not only provides a possible experimental implementation of an active flow network model, but also furnishes a connection between the model parameters and real experimental parameters such as the length and average height of the units. We begin with a description of the active units.

\subsection{Active unit} 

The active unit is described as an hourglass-shaped rigid channel as in Fig.~\ref{Fig_schematic_ring} \textbf{a} (see also Ref. \citenum{Antunes2022}). The inner walls are partially-coated in catalytic material that reacts with the dissolved fuel, leading to the production of the solute. Solute is removed from the system (due to e.g. the inverse chemical reaction, or due to further chemical reactions with a third dissolved species) with a constant rate $\chi$ everywhere in bulk, allowing for steady states to emerge. The solute concentration $\rho(\mathbf{r},t)$ obeys the advection-diffusion equation 
\begin{equation}
\dot{\rho}(\mathbf{r},t)=-\nabla\cdot \mathbf{j}(\mathbf{r},t) - \chi \rho(\mathbf{r},t)\,
\label{eq:adv-diff}
\end{equation}
where $\mathbf{j}(\mathbf{r},t)$ is the flux of the solute. There are three contributions to the flux: one coming from diffusion, one coming from advection due to fluid flow, and the third arises due to the effective interaction between the solute and the channel wall. The flux $\mathbf{j}(\mathbf{r},t)$ is thus given by
\begin{align}
\mathbf{j}(\mathbf{r},t) = &-D\nabla\rho(x,y,t) + \mathbf{v}(\mathbf{r},t)\rho(\mathbf{r},t) -\nonumber \\
&-\beta D\rho(\mathbf{r},t)\nabla U_{wall}(\mathbf{r}),
\label{eq:app-J-0}
\end{align}
where $D$ is the diffusion coefficient of the solute, $\beta$ is the inverse thermal energy, and $U_{wall}(\mathbf{r})$ is the effective wall-solute interaction potential. Equation \eqref{eq:adv-diff} obeys constant flux boundary condition on the channel walls, where solute is produced at a constant rate near the catalyst. We assume the flow velocity $\mathbf{v}(\mathbf{r},t)$ to be small enough for the solution to obey the Stokes equation
\begin{equation}
    0 = - \nabla P(\mathbf{r},t) + \mu \Delta \mathbf{v} - \rho(\mathbf{r},t) \nabla U_{wall}(\mathbf{r}) (\mathbf{r},t), \label{eq:stokes}
\end{equation}
with $\mu$ being the dynamic viscosity and $P(\mathbf{r},t)$ the pressure. The third term encodes the interaction between the solution and the wall and drives the motion of the fluid. We further assume that the fluid is incompressible  
\begin{equation}
    \label{eq:incomp}
    \nabla \cdot \textbf{v}(x,y,t) = 0.
\end{equation}
We construct the active units such that the solute does not escape into the neighboring elastic units, which can be achieved by placing the catalytic patch centered on the bottleneck, far enough away from the borders of the unit such that solute molecules decay before escaping into the elastic units. We note that such a channel will be fore-aft symmetric.

The dynamics of such catalytically-active channels have been previously studied \cite{Antunes2022, Antunes2023, Michelin2015, Michelin2020,Michelin2019,Chen2021}. It has been found that fore-aft symmetric channels can exhibit a spontaneous symmetry breaking that leads to bi-directional pumping \cite{Antunes2022, Antunes2023}. It was further found that asymmetric channels (e.g., due to a ratcheted shape) show hysteresis and discontinuous jumps in the steady-state current. The aforementioned works make use of periodic boundary conditions in the channel's inlets and outlets. We expand on this existing literature by including an applied pressure drop across the channel's ends. Indeed, this set-up is more general, 
since the elastic units to each side of an active unit are generally not symmetric copies of each other. Via lattice Boltzmann simulations (see Supplementary Material \ref{sec:LBM} for details), and an analytical approach based on a combination of lubrication and Fick-Jacobs theories (see Supplementary Material \ref{sec:continuummodel} for the full derivation), we demonstrate that the steady-state current $Q$
\begin{equation}
    Q = \int v_x (\mathbf{r},t) dA
\end{equation}
across the active unit can take either one or two values depending on the magnitude of the applied pressure drop across the ends of the unit (Fig. \ref{Fig_schematic_ring} (c)). 
   

For $\Delta P=0$, the two possible values of the current arises from a spontaneous symmetry breaking where an active unit can either pump to the left or to the right (as seen in Refs. \cite{Antunes2022, Antunes2023}). Even when the symmetry is broken, and there is a pressure drop across the active unit ($\Delta P \neq 0$), there is still bi-stability for low enough values of $|\Delta P|$, as the active flows are still stronger than the passive flows induced by the pressure drop. In such a case, it is still possible to advect solute up the bulk pressure gradient, and thus pump against the pressure drop imposed by the elastic units to each side of the active unit. There is however a critical value of the pressure drop $|\Delta P_c|$ beyond which the active flows cannot push solute up the bulk pressure gradient. In such a case, the active flows can only arise in the same direction as the passive flows induced by the pressure drop. 

In this study, the only information that we need regarding the active unit is the current $Q$, and as such the curve $Q(\Delta P)$ acts as a full description. We take each unit to have the same chemical and geometrical properties, and thus the same curve $Q(\Delta P)$. It must now be said that the pressure drop across a given active unit is not externally imposed, it is read from the pressure field (defined over the whole ring), which needs to be solved also in the elastic unit. The coming sub-section deals with this problem. Furthermore, a discussion of typical experimental parameters for active pores can be found in Section \ref{methods:experimental_values}.

\subsection{Elastic unit}

\label{Method_elastic_units}

The elastic unit is comprised of a thin membrane (made out of homogeneous and isotropic material) which at rest (unstressed) forms a cylinder. This membrane is clamped at both ends, which are connected to active units on each side. The membrane deforms due to the presence of fluids on both sides of it. We take the membrane deformations to be small enough such that the generalized Hooke's law applies~\cite{book_Goncharov}. We further assume that the deformation occurs slowly enough to treat the motion of the membrane as overdamped. As such, the deformation $\mathbf{u}(\mathbf{r},t)$ with respect to the rest configuration is the solution of 
\begin{equation}
    \mu_L \nabla^2 \mathbf{u}(\mathbf{r},t) + (\lambda + \mu_L) \nabla [\nabla \cdot \mathbf{u}(\mathbf{r},t)] = 0,
    \label{eq:deformation}
\end{equation}
where $\mu_L$ and $ \lambda$ are the Lam\'e constants ~\cite{book_Goncharov}.
We take the membrane to move slowly enough such that the pressure field changes adiabatically with $\mathbf{u}(\mathbf{r},t)$, i.e., no sound waves are formed. As such, the fluid inside the elastic unit can still be treated as incompressible. The Stokes equation is similarly valid. Equation \eqref{eq:deformation} must be solved with the boundary condition that the stress tensor must be continuous in the full domain. As such, the deformation profile will depend on the stress tensor of the fluid in the inner side of the membrane, as well as the pressure $\overline{P}$ of the gas on the outer side. The latter is taken to be constant in time and homogeneous in space. 

We assume the flow to be slow enough such that the viscous contribution to the stress tensor can be neglected when compared with the pressure contribution. Furthermore, if the deformations are small enough, the pressure variations inside the elastic unit will be small with respect to the spatially-averaged value $\langle P \rangle (t)$. As such, we approximate the stress tensor in the space surrounded by the elastic membrane as $-\langle P \rangle (t) \delta_{ij}$.

For an elastic unit whose average radius is much smaller than its length, the deformation field away from the clamped ends is well-approximated by the deformation of an infinitely-long cylindrical membrane, which can be obtained analytically \cite{book_Goncharov}. Using this result, we show (see Supplementary Material \ref{sec:continuummodel}) that for a thin membrane, the volume $V(t)$ of the elastic unit obeys 
\begin{equation}
     V(t) - V_0 = \gamma (\langle P \rangle (t) -\overline{P}), \label{eq_VfromP}
\end{equation}
where $V_0=\pi R_1^2 L_e$ is the volume at rest ($\langle P \rangle (t) = \overline{P})$, and $\gamma$ is 
\begin{equation}
    \gamma = \frac{\pi R_1^3}{ 2 \delta R} \frac{2\mu_L + \lambda}{\mu_L(\mu_L + \lambda)} L_e
\end{equation}
to leading order in the membrane thickness $\delta R$. The quantity $R_1$ is the inner radius of the elastic unit at rest and $L_e$ is the length of the elastic unit.
As shown in the Supplementary Material, incompressibility leads to
\begin{equation}
    \partial_t V(t) = Q_{L}(t) - Q_{R}(t), \label{eq:Vdot_fromQ}
\end{equation}
where $Q_{L}(t)$ and $Q_{R}(t)$ are the flow rates coming from the active unit to the left and right of the elastic unit, respectively. Under the framework we just described, the elastic units are entirely represented by two numbers: their volume and their average pressure (which are connected via Eq. \eqref{eq_VfromP}). The volume of an elastic unit evolves in time according to the fluxes imposed on either side from the active units (Eq. \eqref{eq:Vdot_fromQ}). In turn, the pressures of the elastic units (computed from the volumes via Eq. \eqref{eq_VfromP}) determine these fluxes (Fig. \ref{Fig_schematic_ring}). As such, the entire system reduces to a set of differential equations for the pressures and volumes of each elastic unit, which we present (and solve numerically) in Section \ref{sec:discrete}. 

\section{Discrete model of an active flow network with elastic elements}
\label{sec:discrete}

In this section we present and numerically solve the discrete model for the active flow network described in Fig. \ref{Fig_schematic_ring}, and derived in section \ref{sec:methods}.
Our flow network model---represented by Fig.~\ref{Fig_schematic_ring} \textbf{b.2}---is a graph composed of $N$ nodes and $N$ links/connections. The nodes represent the elastic units, while the connections correspond to the active units. For the currents, we use a sign convention such that currents are positive if they flow in the clockwise direction (blue arrows, increasing values of $i$) and negative if they do so in the counter-clockwise direction (red arrows, decreasing $i$). As previously described, the active units can function as micro-pumps capable of sustaining flow against an opposing pressure difference (up to a certain threshold).  To model the behavior of the active unit, we fit the numerical results shown in Fig.~\ref{Fig_schematic_ring} \textbf{c} up to second order in $\Delta P_i$
\begin{equation}
\label{eq:currentDeltaP}
Q_{i}[\Delta P_{i}(t)] =
\begin{dcases}
\begin{aligned}
-\overline{Q}
&+ m\,\Delta P_{i}(t)
- a\,\Delta P_{i}^{2}(t), \\
&\text{if } \Delta P_{i}(t)<\Delta P_{c}
\ \text{and } b(t)=-1,
\end{aligned}
\\[1.5ex]
\begin{aligned}
\ \overline{Q}
&+ m\,\Delta P_{i}(t)
+ a\,\Delta P_{i}^{2}(t), \\
&\text{if } \Delta P_{i}(t)>-\Delta P_{c}
\ \text{and } b(t)=1.
\end{aligned}
\end{dcases}
\end{equation}

where $\Delta P_{i}(t) = P_i(t) - P_{i+1}(t)$ and $b_i(t)$ encodes the branch of the hysteresis the unit currently occupies. The quantity $b_i(t)$ can either take the value of one for the positive (blue) branch, or minus one for the negative (red) branch. In our simulations, when the pressure drop exceeds the critical value $|\Delta P_c|$, the active unit jumps instantaneously to the other branch. In section \ref{methods:experimental_values}, we show that for realistic experimental values, the timescale associated to the switch of branch is indeed negligible. Fig. \ref{Fig_schematic_ring} \textbf{b.1} presents a small section of the ring where currents $Q_{i-1}(t)$ and $Q_{i}(t)$ flow towards elastic unit $i$. Whenever $Q_i \neq Q_{i-1}$ fluid incompressibility will lead to an accumulation (for $Q_i<Q_{i-1}$) or a reduction (for $Q_i>Q_{i-1}$) of the volume ($V_i$)
\begin{equation}\label{Eq_mass_conservation_law}
    \partial_{t}V_{i}(t)=Q_{i-1}[\Delta P_{i-1}(t)]-Q_{i}[\Delta P_i(t)],
\end{equation}
thus the volume of the elastic units can increase or decrease with respect to their rest value $\overline{V}$, leading to a pressure difference with the pressure outside the elastic unit ($\overline{P}$),
\begin{equation}\label{Eq_Local_coupling}
    V_{i}(t)-\overline{V}=\gamma [ P_{i}(t) - \overline{P}]\,,
\end{equation}
where $\gamma$ is the parameter governing the relation (see \ref{Method_elastic_units}). Combining equations \eqref{Eq_mass_conservation_law} and \eqref{Eq_Local_coupling}, and using the characteristic scales $\left[P\right]=\Delta P_{c},\text{ }\left[t\right]=\gamma\Delta P_{c}/\overline{Q},\text{ } \left[x\right]=\Delta x, \text{ }\left[Q\right]=\overline{Q},\text{ } \left[V\right]=\gamma \Delta P_c$, we obtain a dimensionless equation for the time evolution of the pressure field,
\begin{equation}\label{Eq_local_time_evolution_2}
\partial_{\tilde{t}}\tilde{P_{i}}(\tilde{t})=\tilde{Q}_{i-1}[\Delta \tilde{P}_{i-1}(\tilde{t})]-\tilde{Q}_{i}[\Delta \tilde{P}_i(\tilde{t})] \,,
\end{equation}
where variables with a tilde are dimensionless. We integrate these dimensionless equations to get the evolution of the system, but unless stated otherwise we will refer to variables with dimensions along the text. Most of our simulations are initiated with all elastic units in mechanical equilibrium, corresponding to ${V}_{i}(0)=\overline{V}$
and $P_i(0)=\overline{P}, \forall i$. Since then $\Delta P_i(0) = 0$ for all active units, they can pump either in the clockwise ($Q_{i}(0) = \overline{Q}$) or counter-clockwise ($Q_{i}(0) = -\overline{Q}$) direction, i.e. the active units will be in the positive or negative branch ($b_i(0) = \pm 1$), blue and red curves of Fig. \ref{Fig_schematic_ring} \textbf{c}. Next, we integrate the system of differential equations defined in \eqref{Eq_local_time_evolution_2}, to obtain the system's time evolution. We identify two ``trivial'' solutions where all elastic units are in mechanical equilibrium 
($P_i=\overline{P}$
for all $i$ and $t$) and all active units pump in the same direction. As a result, the system sustains a steady current $\pm \overline{Q}$ indefinitely circulating within the ring. However, we also observed a complete different phenomenology presented in the next sections.

\subsection{Spontaneous solitary waves emerge from disorder}

\begin{figure*}
\centering
\includegraphics[width=\linewidth]{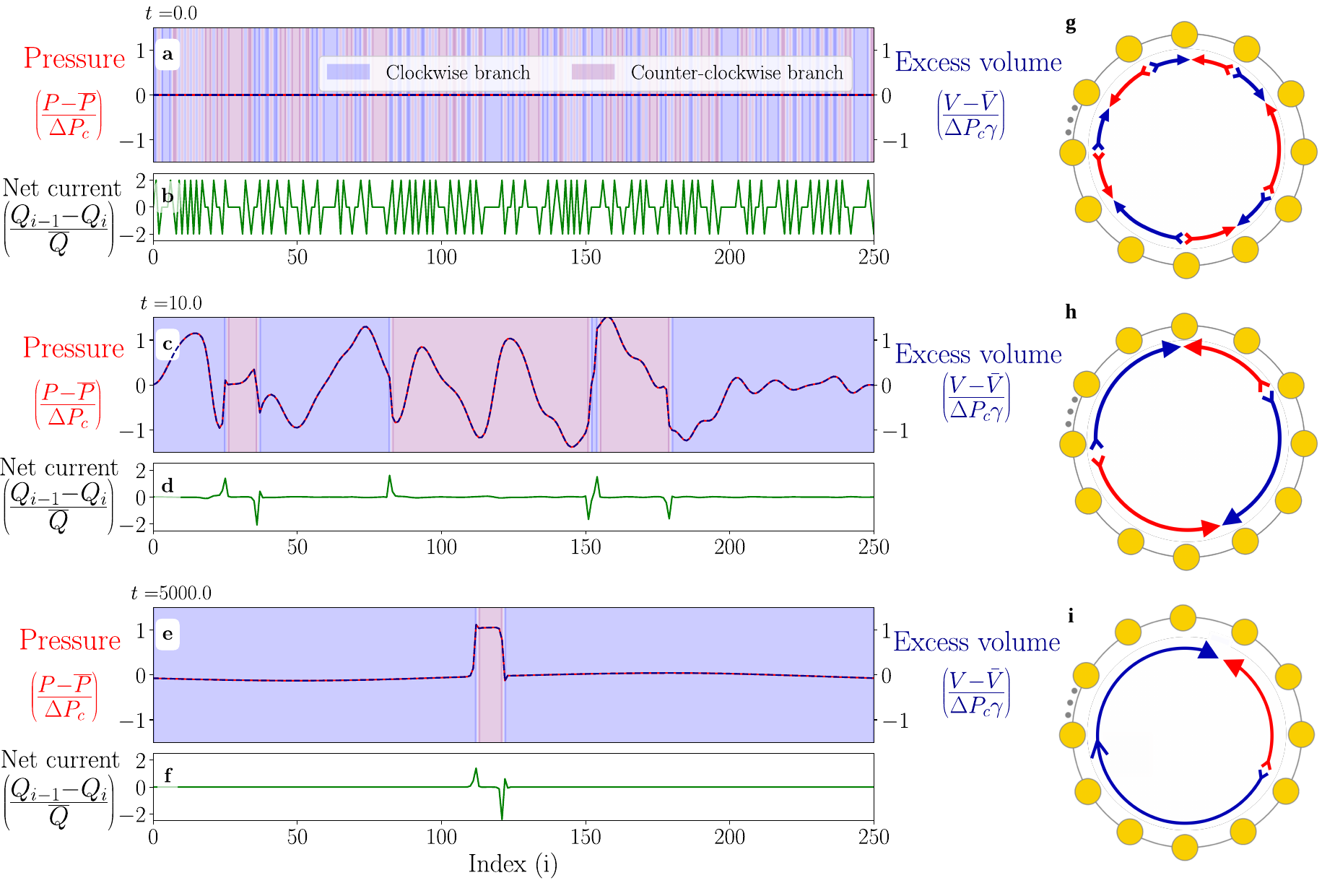}

\caption{Spontaneous emergence of a solitary wave from a disordered configuration. 
The system is initialized at random $b=\pm 1$ and it evolves with time following Eq. \eqref{Eq_local_time_evolution_2}.
Snapshots at $t=0, 10, 5000$ are represented in panels (\textbf{a},\textbf{b}), (\textbf{c},\textbf{d}), and (\textbf{e},\textbf{f}), respectively. Panels \textbf{a}, \textbf{c} and \textbf{e} present pressure (red solid line) and volume (blue dashed line), and the background color displays the branch distribution of the active pores (blue for $b=1$ and red for $b=-1$). Panels \textbf{b}, \textbf{d} and \textbf{f} present the net current (green solid line) at the elastic units.  Each configuration is accompanied by a schematic drawing to help visualization (panels \textbf{g, h, i}), the elastic units are represented by orange circles and the flow in the active units by solid arrows. Simulations are carried out using the dimensionless equations, with relevant parameter values:  $N = 251$, \textcolor{black}{$\overline{Q} \approx 4.61, m \approx 0.117$, $a \approx 2\cdot 10^{-4}$ and $\Delta P_c\approx 23.04$}. }
\label{Fig_soliton_emergence}
\end{figure*}

In Fig. \ref{Fig_soliton_emergence}, we initialize the system in a `disordered' state where $P_i(0)=\overline{P}$ for all elastic units, and $b_i(0) = \pm 1$ is chosen at random for each active unit. Figures \ref{Fig_soliton_emergence} \textbf{a} to \textbf{f} show the time evolution of such a network. At $t=0$ the random distribution of $b_i$ leads to a net current that presents many peaks (Fig. \ref{Fig_soliton_emergence} \textbf{b}). When two adjacent active units are oppositely oriented such that they direct fluid into (away from) the same elastic unit, the accumulated volume in the latter increases (decreases), leading to an  increase (decrease) in pressure. These pressure changes consequentially affect the active units' currents leading to the dynamics displayed in the figure. Ultimately, in Fig. \ref{Fig_soliton_emergence} \textbf{c}, the initially noisy profile smoothens out, making it possible to observe the emergence of large flow domains characterized by their value of $b_i$, in a gradual coarsening process. Smooth oscillations in the pressure field decay following an effective diffusion process (see section \ref{APPENDIX_background_diffusion} of the SM). Finally, as shown in panels \textbf{e, f, i}, for this initial condition, the system reaches a dynamical regime characterized by two well-defined flow domains that propagate at a constant velocity along the ring without changing their profile. This propagation is accompanied by a traveling wave in the pressure/volume fields and localized peaks in the net current at the domain boundaries, giving rise to a solitary wave that we refer to as an Active Solitary Wave (ASW). This solitary wave is an ordered dynamical state of the system spontaneously arising from an initially disordered configuration. A video illustrating this phenomenon is provided in the Supplementary Material \ref{APPENDIX_videos}. Depending on the initial conditions, the system can evolve either to a `trivial' solution with all the active units pumping in the same direction or to a configuration with one or more ASWs. If the pores are initialized at random we find that an ASW appears $\approx 70\%$ of the cases, see \ref{SM_ASW_probability}.

\begin{figure*}
\centering
\includegraphics[width=\linewidth]{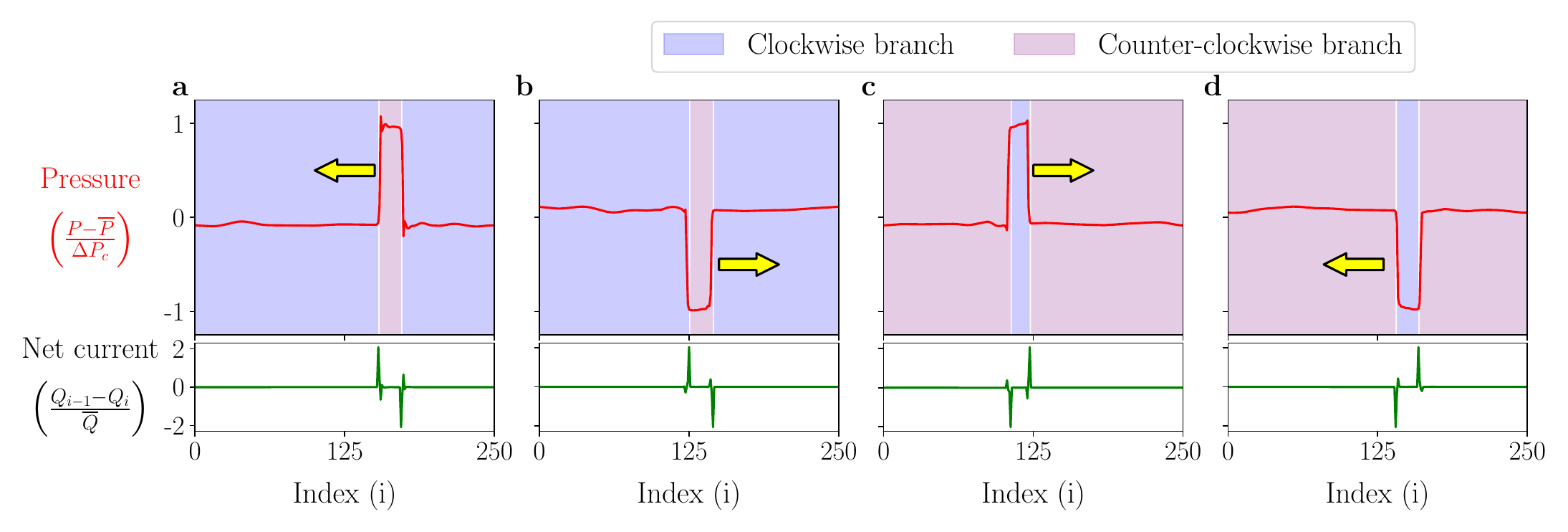}
\caption{Solitary waves can be of four different types. Direction of ASW movement (yellow arrow), shown for different configurations of pressure (red line) and active pore branch domains $b_i$ (background color), blue corresponds to clockwise and red to counter-clockwise. The direction of movement is controlled by the combination of volume accumulation/depletion with the net current at the boundaries of the ASW (solid green line). We initialize the system close to each of these solutions. Simulations are carried out using the dimensionless equations, with relevant parameter values:  $N = 251$, \textcolor{black}{$\overline{Q} \approx 4.61, m \approx 0.117$, $a \approx 2\cdot 10^{-4}$ and $\Delta P_c\approx 23.04$}. }
\label{Fig_movimiento_soliton}
\end{figure*}

We identify four different types of solitary waves, depending on whether the ASW exhibits positive or negative pressure, as well as the sign of the flow within the ASW domain ($b_i=\pm 1$). These four possible combinations appear with equal probability, and are shown in Fig.~\ref{Fig_movimiento_soliton} with the corresponding direction of propagation indicated by a yellow arrow. Each panel follows the same color scheme as in Fig.~\ref{Fig_soliton_emergence}.
The boundaries of the ASW act effectively either as a sink or source (see green line in Fig.~\ref{Fig_soliton_emergence}), leading to the decrease or increase of volume at the elastic units at the boundaries of the ASW (see Fig. \ref{Fig_schematic_ring} \textbf{b}). In the cases where the ASW presents a domain of greater pressure, the `sink' is effectively transferring volume from that boundary to the `source' in the other boundary, making the solitary wave move in the direction of the source (Fig.~\ref{Fig_soliton_emergence} \textbf{a}). The ASW moves one site when the pressure increases at the source (and decreases at the sink) until it reaches $\Delta P_c$ ($0$) and the active units adjacent to the boundary change pumping direction. This is the reason why the ASW has height $ P - \overline{P} \sim \Delta P_c$ in our simulations, and it also determines its velocity (see section \ref{APPENDIX_SECTION_VELOCITY} of the SM). For finite systems, such as the one displayed in the figure, volume conservation leads to $\frac{V_i-\bar{V}}{\gamma \Delta P_c} \ne 0$ away from the ASW. In the cases where the ASW presents a negative pressure domain (volume depletion) the argument is analogous to the one presented above, leading to a motion in the opposite direction. Using the nomenclature of the sine-Gordon paradigm, our ASW is a kink/anti-kink pair that moves synchronously in the same direction. 
Our active flow network can sustain more than one ASW simultaneously. A positive and a negative pressure ASW propagate in opposite directions and inevitably collide, resulting in mutual annihilation (see video at SM \ref{APPENDIX_videos}).

\subsection{Non-local coupling between volume and pressure}

\begin{figure}[!h]
\centering
\includegraphics[width=0.8\linewidth]{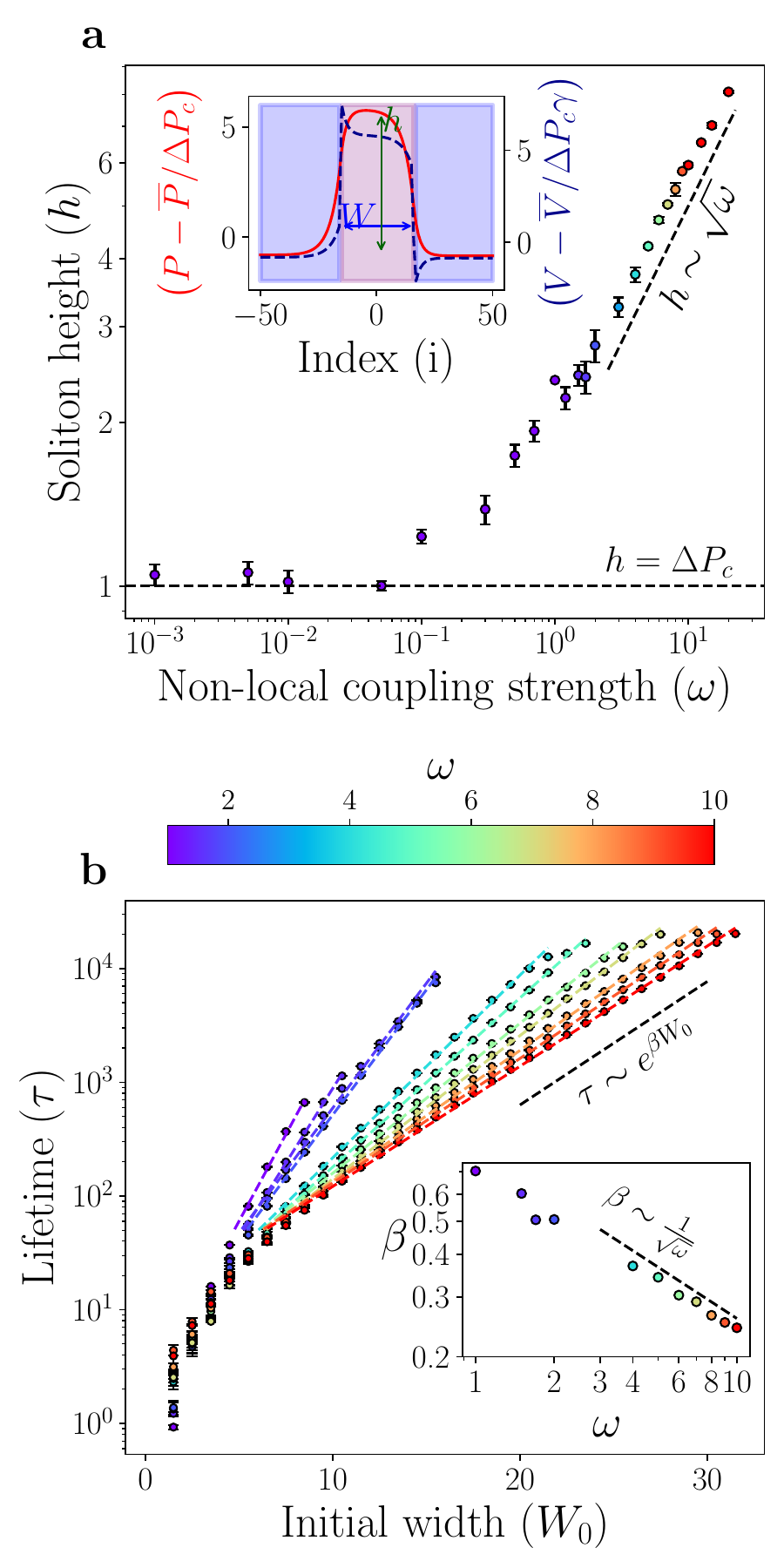}
\caption{Solitary waves for non-local coupling ($\omega \ne 0$). Panel \textbf{a} shows the ASW height as a function of $\omega$. Dashed lines show the theoretical predictions derived in the SM. The inset in this panel illustrates the definitions of the ASW height $h$ (green arrow) and width $W$ (dark blue arrow), as measured from the ASW pressure profile (red solid line) and flow domains (background colors), for a case \textcolor{black}{$\omega = 10$}.  Panel \textbf{b} shows the ASW dimensionless lifetime ($\tau$) from $\omega=1$, blue, to $\omega=10$, red, as a function of the initial width, $W_{0}$.  The data displays a clear exponential behavior for large $W_0$, that we illustrate with a dashed line that follows $\tau \sim e^{\beta W_{0}}$. The inset presents the exponent of the exponential fit ($\beta$) versus $\omega$. In the inset, an analytical prediction for the scaling behavior of $\beta$ is also included as a dashed line (see ~\ref{APPENDIX_sec_SM_Lifetime}). In both panels we initialize the system \textcolor{black}{with a ASW of height $P-\overline{P}=\Delta P_c$}, and the system quickly evolves towards the height presented in panel \textbf{a}, whereas it reduces its width following the behavior shown in panel \textbf{b}. Simulations are carried out using the dimensionless equations, with relevant parameter values:  $N = 251$, \textcolor{black}{$\overline{Q} \approx 4.61, m \approx 0.117$, $a \approx 2\cdot 10^{-4}$ and $\Delta P_c\approx 23.04$. We carried out 5 simulations for each point and the error bars represent the standard deviation of the data}.}
\label{Fig_vida_soliton}
\end{figure}

Until now we have dealt with networks where the coupling between pressure and volume is local, occurring independently at each elastic unit (see equation \eqref{Eq_Local_coupling}). 
However, in several circumstances~\cite{ruiz2021emergent,martinez2024fluidic, altman2025collective,ren2018auxetic,overvelde2015amplifying,lazarus2015soft} the elastic elements cannot be regarded as independent and their coupling may strongly affect the onset of the ASW. In order to account for such a coupling in a simple manner, we add a term proportional to the graph Laplacian ($\mathbf{L}$) of the network. This is the discrete counterpart of the Laplacian operator in the continuum limit,
and the next non-zero term in a graph-based Taylor expansion for the relation between volume and pressure.
Incorporating this new term to  Eq.~\eqref{Eq_Local_coupling} (with dimensions) yields  \begin{equation}\label{Eq_non_Local_coupling} 
     V_{i}(t)-\overline{V}=\sum\limits_{j}\gamma\left( \delta_{ij}+\omega L_{ij}\right)[P_{j}(t)- \overline{P}]  ,
\end{equation}
where $\omega$ is the new parameter that controls the non-local volume-pressure coupling  and $\delta_{ij}$ is the Kronecker's delta ($\delta_{ij}=1$ if $i=j$, and $\delta_{ij}=0$ otherwise). We use the previously defined characteristic scales to render the equations dimensionless 
and the time evolution of the pressure field in the network is given by 

\begin{equation}
\label{Eq_time_evolution_NORM}
\begin{aligned}
&\partial_{\tilde{t}}\tilde{P}_{i}(t)
=
\\
& \sum\limits_{j} \left(
\mathbf{I}
+
\omega \mathbf{L}
\right)_{ij}^{-1} 
\left(
\tilde{Q}_{j-1}[\Delta \tilde{P}_{j-1}(\tilde{t})]
-
\tilde{Q}_{j}[\Delta \tilde{P}_j(\tilde{t})]
\right).
\end{aligned}
\end{equation}

where $\mathbf{I}$ is the identity matrix and once again, tildes over the magnitudes denote dimensionless variables. In the presence of non-local volume–pressure coupling ($\omega \neq 0$), ASWs still emerge from disorder, but their profiles become smoother (see the inset of Fig.~\ref{Fig_vida_soliton} and Fig.~\ref{APENDIX_Fig_soliton_emergence} in the SM). Figure~\ref{Fig_vida_soliton}\textbf{a} shows the ASW height $h$ as a function of $\omega$. The horizontal dashed line at $h = (P - \overline{P}) / \Delta P_c = 1$ indicates the theoretical prediction for the ASW height in the local-coupling limit ($\omega = 0$); as discussed in the previous section, this corresponds to the height required for the active units at the boundaries to flip the sign of the flow. For large $\omega$, the ASW height follows a power-law dependence ($h \sim \omega^{1/2}$), shown by the dashed line in Fig.~\ref{Fig_vida_soliton}\textbf{a}. This scaling can be understood through a simple dimensional argument (see Sec.~\ref{APPENDIX_height_of_soliton} in the SM). In addition to affecting the height, non-locality also modifies the ASW propagation speed, following the same scaling law, $v \sim \omega^{1/2}$ (see Sec.~\ref{APPENDIX_SECTION_VELOCITY}).

When $\omega=0$, the solitary waves remain stable over our longest simulation time, without changing their shape. In contrast, ASWs emerging for $\omega>0$ display a finite lifetime. Figure~\ref{Fig_vida_soliton} \textbf{b} shows the \textcolor{black}{dimensionless} lifetime ($\tau$) of an ASW as a function of their initial width $W_0$ for various values of $\omega$, ranging from $\omega = 1$ (dark blue) to $\omega = 10$ (red).   The lifetime exhibits an exponential dependence on its initial width, following the form $\tau \sim e^{\beta(\omega) W_0}$, represented by dashed lines in the figure.
As observed in the inset of Fig.~\ref{Fig_vida_soliton} \textbf{b}, increasing $\omega$ results in a shorter ASW lifetime, 
captured by $\beta \sim \omega^{-1/2}$. 
This behavior arises from an effective attraction between the ASW boundaries induced by the non-local coupling.
This effective attraction induces a progressive narrowing of the ASW width over time and ultimately leads to its disappearance when the two boundaries collide, see SM for more details (\ref{APPENDIX_sec_SM_Lifetime}) and a video of the phenomenon (\ref{APPENDIX_videos}).

\section{Predictions for experimental realizations}
\label{methods:experimental_values}

Our dimensionless equations facilitate understanding of the system by clearly identifying the combinations of parameters that govern its behavior. However, to motivate the experimental validation of our predictions, it is also useful to consider the characteristic magnitudes of the observed ASW when realistic parameter values are introduced into the model.

Based on the most commonly-used experimental implementation of diffusioosmosis (platinum catalyst in hydrogen peroxide solution), we estimate the active flow velocities to be around $1-10 \upmu m /s  $ \cite{Ebbens2010}.
Based on previous studies on active pores, the bi-stability regime requires an active unit length $L_a \approx 10^2-10^3 \upmu m$ \cite{Antunes2022,Antunes2023}. Taking the height of the active pore to be a tenth of the length, the flow at zero pressure drop $\overline{Q}$ can be estimated to be around $10^3-10^4 \upmu m^3 / s$. The critical value $|\Delta P_c|$ of the pressure beyond which there is only one branch can be estimated by the Poiseuille law, by finding the pressure drop corresponding to a passive flow equal to the active flow we just estimated. Such a calculation (using a dynamic viscosity for aqueous solutions of $\approx 10^{-3} kg/ m/ s $) yields values of $|\Delta P_c|$ between $ 10^{-3} Pa $ and $10^{-1} Pa$. 

For an elastic unit of the length and height ten times that of the active unit, and a membrane of $\approx 100 nm$ thickness built out of PDMS \cite{Pan_2022,Thangawng2007} (with elastic modulus $\approx 1.5 \times 10^6 Pa$ \cite{Huang2018} and Poisson ratio $\approx 0.45$ \cite{Huang2018} determining the Lam\'e coefficients \cite{book_Goncharov} $\lambda \approx 4.7 MPa$ and $\mu_L\approx 0.52 MPa$), the quantity $\gamma$ takes values in the range $\approx 10^{-14} - 10^{-10} m^3/kg/s^2$ leading to a prediction for the ASW velocity around $ L_a \overline{Q}/(\gamma|\Delta P_c|) \approx 10^{-4} m/s$. The derivation of this formula can be found in Supplementary Material \ref{APPENDIX_SECTION_VELOCITY}. Thus, the time it takes for the ASW front to cross an elastic unit is $\approx 10-100 s$. We have assumed the flipping of the active unit to be instantaneous. We may test the quality of this approximation, by comparing the time it takes to flip an active unit with the time to advance the ASW front by one elastic unit. The former is of the same order of magnitude as the time it takes for an initially quiescent flow field to relax to a steady state after a suddenly applied pressure drop. This time is of the order of the diffusive time $L_a^2/\eta_k$, with $\eta_k$ being the kinematic viscosity ($\approx 10^{-6} m^2/s$). As such, the time needed to flip the active unit is $\approx 10^2-10^3$ shorter than the time needed to advance the ASW front, and we thus conclude that the time needed to flip an active unit is negligible.

\section{Conclusion}

In conclusion, the ASWs are reminiscent of topological solitons in the sine-Gordon paradigm; however, they present important differences. The behavior in our system is controlled by two fields (pressure at the nodes and {current} at the edges of the network) that coordinate to produce these emerging structures, allowing the appearance of phenomena that cannot be present in the classic sine-Gordon model  which is described by a single field. In particular, in the sine-Gordon model, kinks and anti-kinks move in opposite directions, and they can pass through each other. In our model, kink and anti-kinks move in the same direction and form pairs that travel without changing their distance (albeit the effective attraction introduced with the non-local coupling between pressure and volume). Even more, if two ASW of different sign collide, they annihilate each other (see video at \ref{APPENDIX_videos}).

Our model has important features in common with other works on active/nonlinear flow networks, which can regulate excitation modes~\cite{forrow2017mode}, mediate information transport~\cite{Woodhouse2018}, and implement logic gates~\cite{Woodhouse2017} in fluidics and, more broadly, in electronics~\cite{kotwal2021active,altman2025collective}. In particular, our model is closer to this previous model introduced for compressible active flow networks~\cite{forrow2017mode}. This work does not describe the emergence of ASWs, although it focused on tree-like networks, which could make their observation difficult. It is particularly suggestive that the authors mention how the dynamics of their model is ``dominated by a significantly reduced number of modes, in contrast to energy equipartition in thermal equilibrium'', which resonates with the observations of Fermi, Pasta, Ulam and Tsingou, who also observed this effect in their famous paper~\cite{fermi1955studies}, what led to the rediscovery of the soliton by Zabusky and Kruskal~\cite{zabusky1965interaction}. We hope that our findings will stimulate further research into the spontaneous emergence of solitary waves in active flow networks with elastic elements.

Finally, our dimensionless equations facilitate understanding of the system by clearly identifying the combinations of parameters that govern its behaviour. However, to motivate the experimental validation of our predictions, it is also useful to consider the characteristic magnitudes of the observed ASW when realistic parameter values are introduced into the model. We have included in section \ref{methods:experimental_values} the typical values for the active and elastic units, leading to a prediction for the ASW velocity of $ \sim 10^{-4} m/s$, where the time it takes for the ASW front to cross an elastic unit is $\sim 10-100 s$. We also estimate that the time needed to flip the active unit is $\approx 10^2-10^3$ shorter than the time needed to advance the ASW front, and thus that the time needed to flip an active unit is negligible (as assumed in our simulations), opening the door to the experimental validation of our results.
We anticipate that these solitary waves could be harnessed in microfluidics and soft robotics to generate complex local flows and enable unconventional modes of information transport, see for example \ref{APPENDIX_SM_open_system} where these ASWs are excited and travel in an open system. This work paves the way for the creation of more advanced fluidic network-based computing \cite{case2020spontaneous,battat2022nonlinear,anandan2015computational,duncan2013pneumatic,preston2019digital}.

\section{Acknowledgements}

We thank Ignacio Pagonabarraga, Miguel Rubí, José Martin Roca and Manuel Mañas for enlightening discussions.
M.R.-G. and R. F.-Q. G. acknowledge support from Ramón y Cajal program (RYC2021-032055-I) funded by MCIN/AEI/10.13039/501100011033 and by European Union NextGenerationEU/PRTR, a Research Grant from HFSP (Ref.-No: RGEC33/2024) with the award DOI (\texttt{\seqsplit{https://doi.org/10.52044/HFSP.RGEC332024.pc.gr.194170}})
and (together with J.J.M.) grant PID2023-147067NB-I00 funded by MCIU/AEI/10.13039/501100011033 and by ERDF/EU.
C.V. acknowledges funding from IHRC22/00002 and Proyecto PID2022-140407NB-C21 funded by MCIN/AEI /10.13039/501100011033 and FEDER, UE. 
M.B. acknowledges funding from the European Research Council under Grant Agreement No. 101117080.
G.C.A., P.M., and J.H. acknowledge funding by the Deutsche Forschungsgemeinschaft (DFG, German Research Foundation) — Project-ID 431791331—SFB 1452.
We acknowledge the use of AI-based language tools (ChatGPT, OpenAI) to assist with the editing and refinement of the manuscript's text.


\setcounter{section}{0} 
\renewcommand{\thesection}{M\arabic{section}} 

\bibliography{references}

\clearpage
\thispagestyle{empty} 
\onecolumngrid

\begin{center}
    \vspace*{3cm} 
    {\LARGE \textbf{Supplementary Materials: Spontaneous emergence of solitary waves in active flow networks}} \\[1.5cm]
    Rodrigo Fernández-Quevedo García, Gonçalo Cruz Antunes, Jens Harting, Holger Stark,\\ 
    Chantal Valeriani, Martin Brandenbourger, Juan José Mazo, Paolo Malgaretti,\\ 
    Miguel Ruiz García \\[0.5cm]
    \today
\end{center}

\clearpage

\setcounter{section}{0} 
\renewcommand{\thesection}{SM\arabic{section}} 

\setcounter{figure}{0}
\renewcommand{\thefigure}{S\arabic{figure}}

\setcounter{equation}{0}
\renewcommand{\theequation}{S\arabic{equation}}

\setcounter{section}{0} 

\renewcommand{\thesection}{SM\arabic{section}} 

\section{Lattice Boltzmann simulations}
\label{sec:LBM}
\begin{figure}
	\centering
   \includegraphics[width=0.5\linewidth]{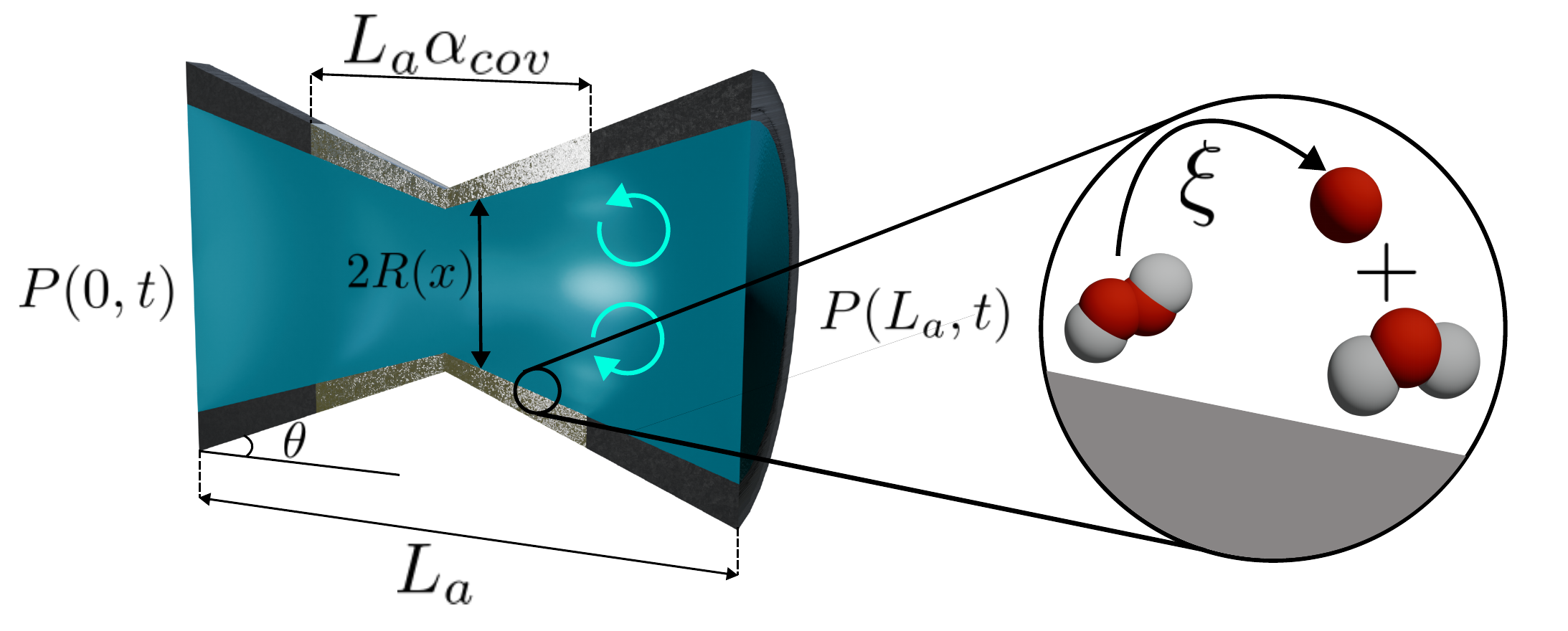}
   \caption{ Diagram describing the model employed in the lattice Boltzmann simulations. The active unit has length $L_a$ and a radius profile $R(x)$ describing an hourglass with inclination angle $\theta$ (Eq. \eqref{eq:SM_hourglass}). The black walls represent chemically inert walls, whereas grey walls are coated in catalytic material. The chemical coating is centered around the bottleneck, extending a distance of $a_{cov}L_a/2$ to each side. Such material leads to the production of a chemical species (solute) with rate $\xi$, as seen in the inset. The presence of a spatially-inhomogeneous concentration of solute leads to diffusioosmotic flows near the wall which drive motion of the bulk fluid as well (light blue arrows). A pressure drop $P_0 - P_L$ is imposed across the pore. 
   \label{fig:LBM_model}}
\end{figure}

 We have performed lattice Boltzmann simulations of {an active unit whose velocity field obeys periodic boundary conditions at the unit's ends. Such} a setup is equivalent to a ring of chained active units (and equivalent to the network discussed in the main text if all elastic units were removed). The simulated model is an adaptation of Refs. \cite{Antunes2022} and \cite{Antunes2023} to include a pressure drop across the active unit. We now present the details of this model.
 
The active unit is shaped like an hourglass (Fig.~\ref{fig:LBM_model}) whose axis of symmetry lies along the $x$-axis. Each unit spans a length $L_a$, has a maximum radius $R_{max}$, and an opening angle $\theta$ defined as the angle the sloping walls make with the $x$-axis (see Fig.~\ref{fig:LBM_model}). The spatially varying radius $R(x)$ is given by
\begin{equation}
\label{eq:SM_hourglass}
R(x) = R_{max}(\theta) - \tan(\theta)\left(\frac{L_a}{2}-\left|x - \frac{L_a}{2}\right|\right).
\end{equation} 
The unit is filled with a Newtonian fluid, the dynamics of which is governed by the continuity equation
\begin{equation}
\label{eq:SM_continuity}
\frac{\partial}{\partial t} \rho_f(\mathbf{r},t) = - \nabla \cdot [\rho_f(\mathbf{r},t) \mathbf{v}(\mathbf{r},t)],
\end{equation}
together with the Navier-Stokes equation 
\begin{equation}
\label{eq:SM_navierstokes}
\rho_f(\mathbf{r},t) \left[ \frac{\partial}{\partial t} \mathbf{v}(\mathbf{r},t) + [\mathbf{v}(\mathbf{r},t) \cdot \nabla] \mathbf{v}(\mathbf{r},t) \right] = - \nabla P(\mathbf{r},t) + \mu \nabla^2 \mathbf{v}(\mathbf{r},t) - \rho(\mathbf{r},t) \nabla U_{wall}(\mathbf{r}) - \rho_f(\mathbf{r},t) \frac{\Delta P}{L_a} \mathbf{e}_x , 
\end{equation}
where $\rho_f(\mathbf{r},t)$ is the fluid mass density, $P(\mathbf{r},t)$ is the pressure, and $\mu$ is the dynamic viscosity. Two body force terms are present in Eq.~\eqref{eq:SM_navierstokes}. The first corresponds to the interaction between solute molecules (with concentration $\rho(\mathbf{r},t)$) and the wall, with an effective potential $U_{wall}(\mathbf{r})$, which shall be defined later. The second body force is a constant force along the $x$ axis which is proportional to the average pressure gradient across the unit. This force is an often-used numerical trick to impose a pressure difference across the unit without imposing a spatially-inhomogeneous fluid density. {Equations~\eqref{eq:SM_continuity} and \eqref{eq:SM_navierstokes} are not solved directly. Instead, we solve a discretized form of the Boltzmann equation (with the Bhatnagar–Gross–Krook collision operator). It can be shown via a Chapman-Enskog analysis that such a procedure yields the Navier-Stokes equation in the limit of low Mach number \cite{Krueger_book}. In such limits,} the fluid density $\rho_f$ is approximately homogeneous, and
\begin{equation}
\label{eq:SM_continuity2}
\nabla \cdot \mathbf{v}(\mathbf{r},t) \approx 0.
\end{equation}
{Furthermore, we perform all simulations in the limit of low Reynolds number such that}
\begin{equation}
\label{eq:SM_navierstokes2}
- \nabla P(\mathbf{r},t) + \mu \nabla^2 \mathbf{v}(\mathbf{r},t) - \rho(\mathbf{r},t) \nabla U_{wall}(\mathbf{r}) - \rho_f \frac{\Delta P}{L_a} \mathbf{e}_x \approx 0. 
\end{equation}
We now require boundary conditions for the velocity field: at the unit walls, no-slip boundary conditions are enforced, while at the unit open ends at $x=0$ and $x=L_a$, periodic boundary conditions are used instead. 

The unit's walls are patterned with a catalytic coating in the section $x \in \{L_a/2- a_{cov}L_a/2, L_a/2 + a_{cov}L_a/2\}$, where the covering fraction $a_{cov}$ can vary from zero (no coating) to one (full coating). This catalytic material enables a chemical reaction in the fluid resulting in the local synthesis of solute. As in the main text, we consider the fuel species that is consumed by the chemical reaction to be so numerous that it never gets fully exhausted. We also consider the consumption of fuel to be reaction-limited, and so the fuel concentration is again homogeneous in space. We consider the solute to be decomposed homogeneously in the bulk fluid with rate $\chi$ (with dimension sec $^{-1}$). Such a sink term may represent e.g. the inverse chemical reaction that consumes solute and produces fuel.

The effective interaction potential $U_{wall}(\mathbf{r})$ between the solute molecules and the active unit walls is taken to be a simple piece-wise linear function of the distance $d(\mathbf{r})$ from the wall,
\begin{equation}
\label{eq:SM_potentialMain}
U_{wall}[d(\mathbf{r})] = \begin{cases} 
      U_0[1 -d(\mathbf{r})/l] & 0\leq d(\mathbf{r})\leq l, \\
      0 & l\leq d(\mathbf{r}),
   \end{cases}
\end{equation}
where $l$ is the range of $U_{wall}$. It is assumed to be much smaller than the average radius of the pore $R_a = R_{max} - \tan(\theta)L_a$. The solute concentration is given as in the main text
\begin{equation}
	\label{eq:SM_rho}
	\dot{\rho(\mathbf{r},t)} = - \nabla \mathbf{j}(\mathbf{r},t) - \chi \rho(\mathbf{r},t),\quad 
\mathbf{j}(\mathbf{r},t)= -D\nabla\rho(\mathbf{r},t) -\beta D\rho(\mathbf{r},t)\nabla U_{wall}(\mathbf{r})+ \mathbf{v}\rho(\mathbf{r},t),
\end{equation}
where $D$ is the diffusion coefficient of the solute, $\beta=1/(k_BT)$ is the inverse thermal energy, and $\mathbf{j}(\mathbf{r},t)$ is the solute flux. Note that the term involving $\chi$ acts as the sink of solute, while the solute source is enforced via the boundary conditions. Equation~\eqref{eq:SM_rho} obeys periodic boundary conditions on the open ends of the unit, and flux boundary conditions on the unit walls, 
\begin{equation}\label{eq:SM_source_sims}
\bf{j} \cdot \bf{n} |_{\text{(x, $\phi$, $y^2 + z^2=R^2(x)$)}} = \begin{cases} 
      \xi, &  |x-\frac{L_a}{2}| < a_{cov}\frac{L_a}{2}, \\
      0, &  \text{otherwise},
   \end{cases}
\end{equation}
where $\phi$ is the azimuthal angle, $\mathbf{n}(x,\phi)$ is a unit vector perpendicular to the pore wall (pointing towards the inside of pore), and $\xi$ is a positive constant with dimension $[m^2 s]^{-1}$.

The wall-solute interaction results in a laterally inhomogeneous pressure along the wall, hence coupling Eq.~\eqref{eq:SM_rho} with Eqs. \eqref{eq:SM_continuity2} and \eqref{eq:SM_navierstokes2}. These three equations are solved in parallel using a finite-difference solver for the first one (second-order in space and first-order in time), and the lattice Boltzmann method (LBM)~\cite{Benzi1992,Krueger_book,Harting2016} for the other two ones. Some details of the numerical implementation can be found in Ref.~\cite{Peter2020}. To adimensionalize our results, we introduce the length $L_a$ as a length scale (corresponding to 40 spatial lattice units), and $\tau_f = L_a^2/\nu$ (9600 temporal lattice units) as a time scale. Note that the latter is the time associated to viscous transport of linear momentum across the unit.

\begin{figure}[t]
	\centering
   \includegraphics[width=0.7\linewidth]{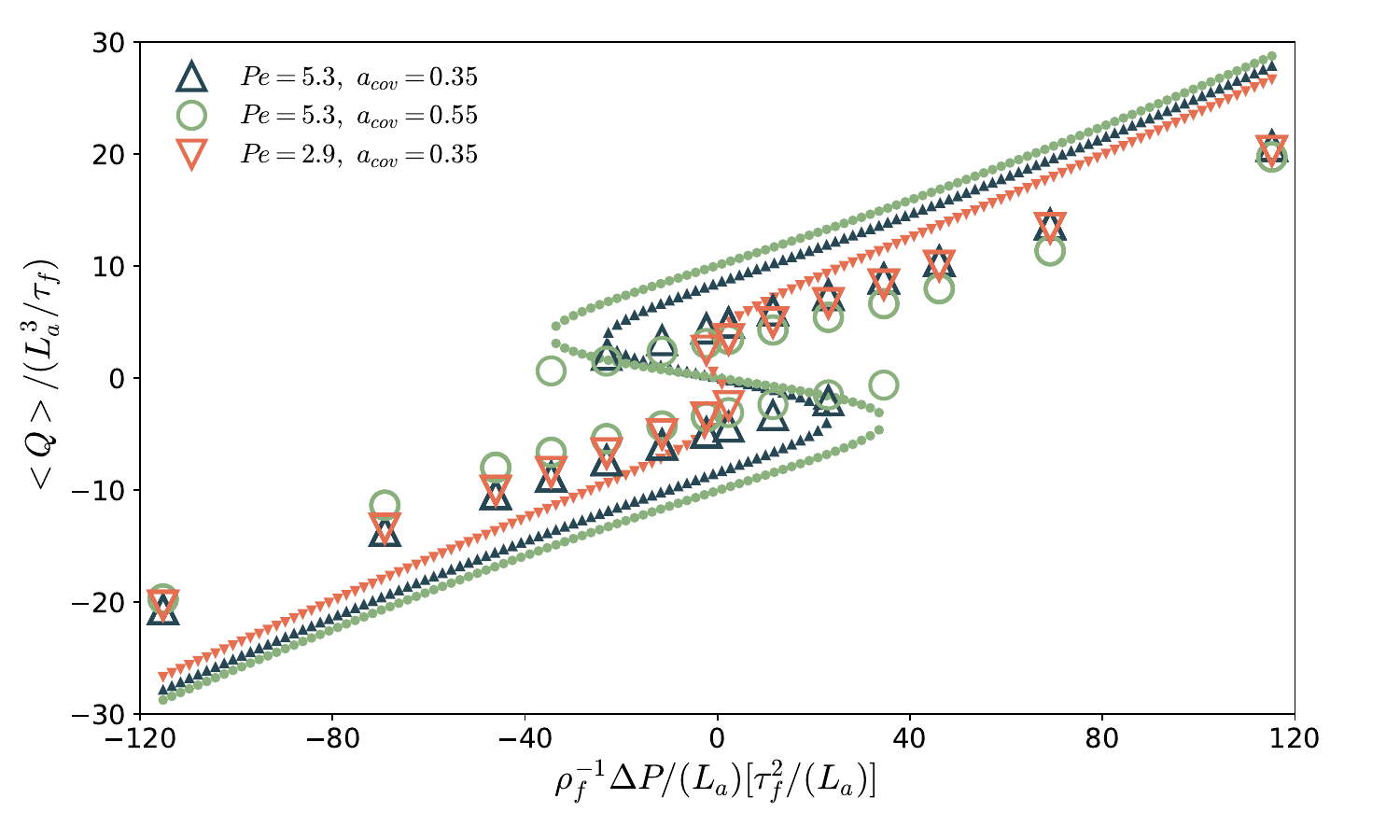}
   \caption{Normalized time-averaged flow rates $\langle \tilde{Q}(t) \rangle$ as a function of normalized pressure drop $\Delta P$. Open points are obtained from lattice Boltzmann simulations: in all three cases, $\theta = \pi/6$, $R_{max}/L_a=1$, $\nu \tau_f/L_a^2= 1$, $\beta U_0 = 4 \times 10^ {-4}$, $d/(L_a)=0.1$, $\xi (L_a)^2\tau_f=1.5 \times 10^7$, and $\chi \tau_f = 9.6$. In lattice units: $L_a=40$, $\eta = 1/6$, $U_0 = 4 \times 10^ {-4}$, $d=4$, $\xi=1$, $\chi=10^{-3}$, $\beta=1$. The simulation box is of size $80\times80\times40$. The filled-in points are obtained from the lubrication theory of Supplementary Material \ref{sec:continuummodel} upon a fitting of the solute production rate, such that $\xi_{lub} = \xi_{LBM} \alpha_{fit}$ where $\alpha_{fit} = \{259 (\bigtriangleup), 212 (\bigcirc ), 176 (\bigtriangledown)$. Furthermore, $(\mathcal{L}/\beta\eta) (\tau_{\chi}/L_a^5) = -6.25 \times 10^{-8}$. 
   \label{fig:QdeltaP}}
\end{figure}

It is known (see Refs. \cite{Antunes2022, Antunes2023}) that the model described above shows a spontaneous symmetry breaking for high enough values of $Pe$, defined as $v^* L_a/(2D)$, where $v^* = 1.3 L_a / \tau_f$ is a typical velocity scale, deemed to be appropriate for the typical parameter values used here \cite{Antunes2022}. We now perform simulations for high enough values of $Pe$ such as to induce this spontaneous symmetry breaking which allows an active unit to pump in either direction. We introduce an externally-imposed pressure drop via the body force in Eq. \eqref{eq:SM_navierstokes2} and measure the value of the (normalized) volumetric flow rate{/current} across the unit  
\begin{align}
   \tilde{Q}(t) = \frac{\tau_f}{(L_a)^3} \int_{0}^{R(x)} dr \ r  \int_0^{2\pi}d \phi \ v_x(r,\phi,x,t), 
\end{align}
where $r = \sqrt{x^2 + z^2}$. For intermediate valuess of $Pe$, this quantity reaches a steady state value, whether as for large values of $Pe$, a pulsating flow state emerges, and $\tilde{Q}(t)$ displays sustained oscillations \cite{Antunes2022}. For both regimes, we time-average $Q(t)$ after a period of initial relaxation, and over a time interval that is much larger than the typical period of the sustained oscillations. This value $\langle Q(t) \rangle$ is reported as a function of the applied pressure drop $\Delta P$ in Fig. \ref{fig:QdeltaP} for three different parameter sets where $Pe$ and $a_{cov}$ is varied. {Note that this Figure contains the same data as Fig. \ref{Fig_schematic_ring} d, except that the units for current and pressure drop are now included explicitly. }In the bistable region, the two branches are accessible by varying the initial condition of the simulation. In all cases, the curves $\langle Q(t) \rangle (\Delta P)$ display bistability and hysteresis. We further plot the results from the semi-analytical theory in section \ref{sec:continuummodel}. In this theory, diffusioosmosis is captured via an effective slip velocity that is proportional to the gradient of solute concentration along the wall (Eq. \eqref{eq:SM_vslip}. As per Refs. \citenum{Antunes2022} and \citenum{Antunes2023}, this proportionality constant was taken to be $(\mathcal{L}/\beta\eta) (\tau_{\chi}/L_a^5) = -6.25 \times 10^{-8}$. The simulations were performed away from the theory's regime of validity, as the former shows aspect ratios close to one ($R_0/L_a = 0.85$). To compensate, we fit the maximum value of the source function of the theory $\xi_{sims}$ so as the minimize the difference between the theory and the simulations in the bistable regime. Both then showcase qualitatively similar curves of $Q(\Delta P)$.

\section{Full treatment of the continuum model for the active elastic network}
\label{sec:continuummodel}


\begin{figure*}
	\centering
   \includegraphics[width=\textwidth]{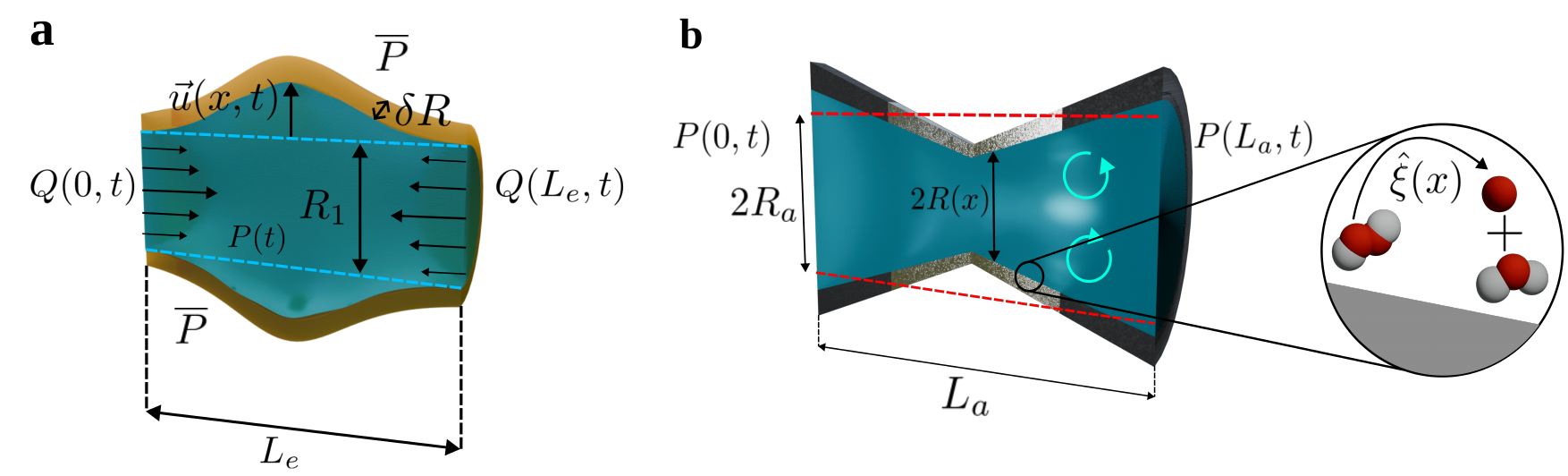}
   \caption{ Panel a: Diagram describing the elastic unit of length $L_e$, membrane thickness $\Delta R$, and an interior radius $R_1$ at rest. The imbalance of pressure inside $P$ and outside $\overline{P}$ the unit leads to a deformation field $\mathbf{u}$. Furthermore, the active units to each side impose a flow on the boundaries of the elastic unit. Panel b: Diagram describing the active unit of length $L_a$, with a varying radius $R(x)$ whose average value is $R_a$. The walls of the active unit are inhomogeneously coated in a catalytic material which leads to the production of solute with a rate $\hat{\xi}$ as seen in the inset. The inhomogeneity in chemical composition thus generated leads to diffusioosmotic flows (light blue arrows). Furthermore, the elastic units to each side of the active unit may be at different pressures.
    \label{fig:elastic}}
\end{figure*}

In this section, we present a model for an active flow network coupled to elastic reservoirs and coarse-grain it into a lattice model that can be easily solved computationally. The approach is based on linear elastostatics \cite{book_Goncharov}, the lubrication theory for viscous flow in narrow channels \cite{Schlichting1979, Hori2006, Antunes2024}, coupled with the Fick-Jacobs theory for diffusive transport also in narrow channels \cite{Zwanzig1992, Burada2007, Reguera2001, Malgaretti2023}. The latter two have been combined before to study catalytically-active pores/channels \cite{Antunes2022, Antunes2023} and active films \cite{Richter2025}, as well as the flow of particles suspended in electrolytes \cite{Malgaretti2014}, and transport in oscillating channels \cite{Marbach2018}. The result is an effective one-dimensional theory which is typically much easier to handle. We begin by presenting the governing equation in section \ref{sec:SI_governingEquations}, moving on to discuss the elastic and active units in more detail in sections \ref{sec:SI_elastic} and \ref{sec:SI_active}, respectively. Finally, in section \ref{sec:SI_lattice}, the coarse-graining into a lattice model is shown.

\subsection{Governing equations}
\label{sec:SI_governingEquations}
Our goal is to study fluid flow inside a chain of units of two distinct classes (elastic and active). While the two kinds of units provide different boundary conditions to the fluid, the equations that rule it are the same in the entire system. Furthermore, both kinds of units can be described as a tube of a certain length, whose axis of symmetry lies along the $x$-axis, and has a spatially-varying radius $R(x,t)$. Because the tube is radially-symmetric, the velocity field $\mathbf{v}(\mathbf{r},t)$ will show the same symmetry. We take $R(x,t)$ and the flow velocity $\mathbf{v}(\mathbf{r},t)$ to be small enough so that the fluid obeys the Stokes equation
\begin{equation}
    0 = - \nabla P(x,t) + \mu \Delta \mathbf{v}(\mathbf{r},t), \label{eq:SM_stokes}
\end{equation}
where $\mu$ is the viscosity and $P(x,t)$ is the pressure. We assume any variation in time of the tube's shape to be slow enough so that the pressure field changes adiabatically with $R(x,t)$, i.e., no sound waves are formed. As such, the liquid can be treated as incompressible
\begin{equation}
    \label{eq:SM_incomp}
    \nabla \cdot \mathbf{v}(\mathbf{r},t) = 0,
\end{equation}
where $\mathbf{v}(\mathbf{r},t)$ is the fluid velocity, which obeys the following boundary conditions at the unit walls
\begin{equation}
    \mathbf{v}(x,r=\pm R(x,t), \phi,t) = \mathbf{v}_w(x,\phi,t) + \mathbf{v}_a(x,\phi,t),
    \label{eq:SM_v_bc}
\end{equation}
where $r$ is the distance to the $x$-axis, and $\phi$ is the azimuthal angle. There are two contributions, the former being the velocity of the tube wall itself $\mathbf{v}_w(x,\phi,t)$ (which is zero in the active unit), and the latter a slip velocity $\mathbf{v}_a(x,\phi,t)$ encoding the diffusioosmotic flows near the wall due to the spatially-inhomogeneous concentration of solute \cite{Anderson1989}. The latter will be zero in the elastic unit. The slip velocity obeys
\begin{equation}
\label{eq:SM_vslip}
\mathbf{v}_a(x,\phi,t)=-\frac{\curlyL}{\beta\mu}\nabla_{||}\rho(x,r = \pm R(x,t), \phi, t),
\end{equation}
where $\beta = 1/(k_BT)$ is the inverse thermal energy and $\rho(\mathbf{r},t)$ is the concentration of solute \cite{Anderson1989}. Furthermore, $\nabla_{||}$ is the derivative along the wall. The phoretic mobility $\curlyL$ is given by \cite{Anderson1989}
\begin{equation}
\curlyL = \int\limits_0^{\infty} r^{\prime}\left\{\exp\left[ -\beta U_{wall}(r^{\prime}) \right] -1  \right\} dr^{\prime}, \label{eq:SM_curlyL}
\end{equation}
where $r^{\prime}$ is the distance to the wall and $U_{wall}(r^{\prime})$ is the effective interaction potential between the solute particles and the wall. This expression for the phoretic mobility assumes a potential $U_{wall}$ that depends only on the distance to the wall, and a radius of curvature of the wall that is much larger than the potential range\cite{Anderson1989}. Additionally, the solute concentration is taken to be in thermal equilibrium within this thin layer where $U_{wall} \neq 0$ \cite{Anderson1989}. The governing equation for the solute concentration in the full domain shall be presented later.

For the elastic unit, the tube is an elastic membrane with a given thickness and shape at rest (vacuum) as depicted in Fig. \ref{fig:elastic}. The presence of fluids on both sides of this membrane lead to a deformation $\mathbf{u}(\mathbf{r},t)$ which we take to be small enough such that the generalized Hooke's law \cite{book_Goncharov} applies and, 
\begin{equation}
      \sigma_{ik}(\mathbf{r},t) = \lambda e_{jj}(\mathbf{r},t) \delta_{ik} + 2 \mu_L e_{ik}(\mathbf{r},t), \label{eq:Hooke}
\end{equation}
where $\boldsymbol{\sigma}(\mathbf{r},t)$ is the stress tensor, $\lambda$ and $\mu_L$ are the Lam\'e constants of the material, and $\mathbf{e}(\mathbf{r},t)$ is the strain tensor
\begin{equation}
    e_{ik}(\mathbf{r},t) = \frac{1}{2} [ \partial_k u_i(\mathbf{r},t) + \partial_i u_k(\mathbf{r},t)].\label{eq:strain}
\end{equation}
We take the motion of the elastic material to be slow enough that the inertial term can be neglected
\begin{equation}
    \partial_k \sigma_{ik}(\mathbf{r},t) = 0, \label{eq:eq_membrane}
\end{equation}
yielding
\begin{equation}
    \mu_L \nabla^2 \mathbf{u}(\mathbf{r},t) + (\lambda + \mu_L) \nabla [\nabla \cdot \mathbf{u}(\mathbf{r},t)] = 0, \label{eq:deform}
\end{equation}
which is to be solved under the condition that the stress tensor is continuous in the boundaries between fluids and membrane. 

 We now describe the equations that govern the solute concentration. The active unit's inner walls are partially-coated in catalytic material that reacts with a given dissolved chemical compound (named the fuel), leading to the production of another dissolved chemical compound (named the solute). In typical experimental implementations, the catalyst is platinum, the fuel is hydrogen peroxide, and the solute is oxygen \cite{Ebbens_Review, Juliane_Review, Howse2007}. We account for removal of solute due to either the inverse chemical reaction, or due to reaction with a third chemical species dissolved in the liquid. This solute removal occurs with a homogeneous rate $\chi$ in the bulk. We consider the solute production to be reaction-limited and the fuel to be so abundant that solute production rates can be considered constant in time. The solute concentration obeys the advection-diffusion equation 
\begin{equation}
\dot{\rho}(\mathbf{r},t)=-\nabla\cdot \mathbf{j}(\mathbf{r},t) - \chi \rho (\mathbf{r},t)\,
\label{eq:SM_adv-diff}
\end{equation}
where $\mathbf{j}(\mathbf{r},t)$ is the flux of solute. There are two contributions to the flux: one coming from diffusion, and one coming from advection due to fluid flow. Thus, the flux $\mathbf{j}(\mathbf{r},t)$ is given by
\begin{equation}
\mathbf{j}(\mathbf{r},t) = -D\nabla\rho(\mathbf{r},t) + \mathbf{v}(\mathbf{r},t)\rho(\mathbf{r},t),
\label{eq:SM_app-J-0}
\end{equation}
where $D$ is the diffusion coefficient of solute. Note that we do not include the drift due to the gradient of the interaction potential $U_{wall}$, as the potential's range is much smaller than the typical channel height. The potential's role is then only to trigger diffusioosmosis and an effective non-zero slip velocity. We have the following boundary conditions on the tube walls:
\begin{equation}
    \mathbf{j}(x,r = \pm R(x,t), \phi ,t) \cdot \mathbf{n} (x,r = \pm R(x), t) = - \hat{\xi}(x) \label{eq:SM_bc_rho}
\end{equation}
where $\hat{\xi}(x)$ encodes the production of solute due to the chemical reaction and depends on the position, as does the catalytic coating. For example, in the elastic-walled elements, there is no catalyst and $\hat{\xi}=0$. 

We perform an additional simplification relating to the solute concentration. We assume the solute production rate to be localized near enough the center of the active units, and the removal of solute in bulk to be intense enough that only a negligible amount of solute can be found in the elastic units.

We will now look at a network composed of $N$ active units and $N$ elastic units, joined together into a chain alternating such that each active unit has two elastic units as neighbours and viceversa. The units on the edge of this chain are joined together to form a ring. The radius of the ring is assumed to be much larger than the typical local radii of each unit, and so any effects arising from curvature may be neglected. The fact that active units do not exhibit elastic behaviour, and that elastic units do not exhibit active slip is the basis of the coarse-graining into a lattice model, as the two kinds of elements are effectively decoupled, barring from the contact points where the pressure $P(x,t)$ and the velocity field $\mathbf{v}(\mathbf{r},t)$ must remain continuous. As such, we now solve the governing equations in both active and elastic units separately. We do so approximately, by employing lubrication and Fick-Jacobs theory, which are valid when the typical local radius of a unit is much smaller than its length.

\subsection{Elastic units}\label{sec:SI_elastic}
We will first concern ourselves with the elastic units. The results we shall derive do not depend on the exact element we examine and as such, to lighten the notation, we change our frame of reference such that $x=0$ is located at the leftmost point of any given elastic unit. We begin by determining the deformation of the membrane via Eq. \eqref{eq:deform}. This equation is to be solved with two sets of boundary conditions: the stress tensor must be continuous at $r=R_1$ and $r=R_1 + \delta R$; the deformation $\mathbf{u}(x,r,\phi,t)$ must be zero at the edges ($x=0$ and $x=L_e$, as the membrane is clamped. The stress tensor outside of the elastic unit ($r > R_1 +\delta R$) is simply that of a still fluid at pressure $\overline{P}$
\begin{equation}
    \sigma_{ik}(x,r>R_1 + \delta R,t) = -\overline{P} \delta_{ik},  \label{eq:stresstensor0}
\end{equation}
while inside the unit, we have a viscous contribution due to the motion of the liquid
\begin{equation}
    \sigma_{ik}(x,r < R_1 ,t) = -P(x,t) \delta_{ik} + \mu \left( \partial_i v_j + \partial_j v_i  \right). 
\end{equation}
We assume the motion of the fluid to be slow enough such that the viscous drag on the membrane may be ignored and so
\begin{equation}
    \sigma_{ik}(x,r = R_1 ,t) = -P(x,t) \delta_{ik}. \label{eq:stresstensor1}
\end{equation}
Furthermore, if the deformation of the membrane is small enough, the pressure will not vary strongly inside the tube and
\begin{equation}
    P(x,t) \approx \langle P \rangle (t),
\end{equation}
where $\langle P \rangle (t)$ is the spatial average of the pressure inside the elastic unit. In such a regime, Eq. \eqref{eq:stresstensor1} yields
\begin{equation}
    \sigma_{ik}(x,r = R_1 ,t) = -\langle P \rangle (t) \delta_{ik}.  \label{eq:stresstensor2}
\end{equation}

Furthermore, if the length of the elastic unit $L_e$ is much larger than the thickness of the membrane and the deformation is small, the effect of clamping the boundaries at the edges will only be noticed close to these edges. For most of the tube, the deformation field of the elastic membrane will be well-approximated by that of an infinite cylinder. The solution of Eq. \eqref{eq:deform} for an infinitely-long cylindrical membrane with boundary conditions given by Eqs. \eqref{eq:stresstensor0} and \eqref{eq:stresstensor2} is given in Ref. \citenum{book_Goncharov} as
\begin{equation}
    \mathbf{u}(r,t) = \left[ \frac{R_1^2 \langle P \rangle(t) - (R_1 + \delta R)^2 \overline{P}}{2 (\lambda + \mu_L)(2R_1 \delta R + \delta R^2)}r  + \frac{\langle P \rangle(t) - \overline{P}}{2\mu_L}\frac{R_1^2(R_1 + \delta R)^2}{2R_1 \delta R + \delta R^2}\right] \mathbf{e}_r. \label{eq:deformSol}
\end{equation}
For thin membranes 
\begin{equation}
    \frac{\delta R}{R_1} \ll 1,
\end{equation}
and the right-hand-side of Eq. \eqref{eq:deformSol} is well-approximated by the leading order term of its expansion in powers of $\delta R/R_1$, yielding
\begin{equation}
     \mathbf{u}(r,t) = \frac{R_1^2}{4 \delta R} \frac{2 \mu_L + \lambda}{\mu_L(\mu_L + \lambda)}(\langle P \rangle(t) - \overline{P}) \mathbf{e}_r.
     \label{eq:SM_displacement_final}
\end{equation}

Having determined how the elastic membrane deforms, we describe the motion of the fluid inside of it. We take the incompressibility condition of Eq. \eqref{eq:SM_incomp}, which we integrate over the cross-section, yielding
\begin{equation}
    \int\limits_{0}^{R(x,t)} r \partial_x v_x (x,r,t) dr = -R(x,t) v_r [x,R(x,t),t], 
\end{equation}
or
\begin{equation}
   \partial_x  \left[ \int\limits_{0}^{R(x,t)} r v_x (x,r,t) dr \right] - R(x,t) v_x[x,R(x,t),t] \partial_x R(x,t) = -R(x,t) v_r [x,R(x,t),t]. 
\end{equation}
which can be written as
\begin{equation}
    \partial_x Q(x,t) = 2\pi R(x,t) \left(v_x[x,R(x,t),t] , v_r [x,R(x,t),t] \right) \cdot \left( \partial_x R(x,t), -1\right), \label{eq:SM_getQ}
\end{equation}
where $Q(x,t)$ is the volumetric flow rate{/current}
\begin{equation}
    Q(x,t) = 2\pi\int\limits_{0}^{R(x,t)} r v_x (x,r,t) dr.
\end{equation}
We identify the last vector in Eq.\eqref{eq:SM_getQ} as a non-unitary outer normal to the curve $y = R(x,t)$. Due to the system's axial symmetry, we may then write
\begin{equation}
    \partial_x Q(x,t) = -2 \pi R(x,t) \sqrt{ 1 + [\partial_x R(x,t) ]^2 } \mathbf{v}[x,R(x,t), \phi,t] \cdot \mathbf{n}(x,\phi,t). \label{eq:SM_dxQ_0}
\end{equation}
where $\mathbf{n}(x,\phi,t)$ is the unitary outer normal. We now restrict our analysis to thin unit, such that the typical average radius $R_e$ is much smaller than the length $L_e$
\begin{equation}
    \frac{R_e}{L_e} \ll 1.
\end{equation}
Under this regime, we neglect any terms in Eq. \eqref{eq:SM_dxQ_0} (and future equations in this sections) which are quadratic or above in $\partial_x R(x,t)$ (which is $\ll 1$), yielding
\begin{equation}
    \label{eq:SM_dxQ}
    \partial_x Q(x,t) = -2 \pi R(x,t)  \mathbf{v}[x,R(x,t),\phi,t] \cdot \mathbf{n}(x,\phi,t),
\end{equation}
which using Eq. \eqref{eq:SM_v_bc} reduces to 
\begin{equation}
    \label{eq:SM_dx_Qvw}
    \partial_x Q(x,t) = -2 \pi R(x,t)  \mathbf{v}_w(x,\phi,t) \cdot \mathbf{n}(x,\phi,t),
\end{equation}
Because the tube deforms in response to a pressure imbalance between the inner and outer side of the membrane, and because this pressure imbalance leads to a force that is perpendicular to the membrane, we neglect any tangential motion
\begin{equation}
    [\mathbf{I} - \mathbf{n}(x,\phi,t) \otimes \mathbf{n}(x,\phi,t)]\mathbf{v}_w[x, \phi, t]  = 0,
\end{equation}
yielding
\begin{equation}
    \label{eq:SM_v_wx}
    v_{w,x}(x,t) = - v_{w,r}(x,t) \partial_x R(x,t),
\end{equation}
where $ v_{w,x}(x,t)$ and $v_{w,r}(x,t)$ are the longitudinal and radial components of $\mathbf{v}_w(x,\phi,t)$. Plugging Eq. \eqref{eq:SM_v_wx} into Eq. \eqref{eq:SM_dxQ} yields  
\begin{equation}
    \label{eq:SM_dxQ2}
    \partial_x Q(x,t) = -2\pi R(x,t)  v_{w,r}(x,t).
\end{equation}
The variation of the radius profile $R(x,t)$ in time depends on $\mathbf{v}_{w}(x,t,\phi)$. Given an infinitesimal displacement (over a time $\delta t$) of the tube's inner surface $\mathbf{v}_{w}(x,t,\phi) \delta t$, the local radius changes by a quantity $\mathbf{v}_{w}(x,t,\phi) \cdot \mathbf{n}(x,\phi,t) / \cos (\theta) \delta t$, where $\theta$ is the angle that the curve $y = R(x,t)$ makes with the $x$-axis. Knowing that $\tan(\theta) = \partial_x R(x,t)$, we obtain to linear order in $\partial_x R(x,t)$
\begin{equation}
\label{eq:SM_h_as_vel}
    \partial_t R(x,t) = v_{w,r}(x,t)
\end{equation}
which plugged into Eq.\eqref{eq:SM_dxQ2} leads to
\begin{equation}
    \partial_x Q(x,t) = -2  \pi R(x,t) \partial_t R(x,t),
\end{equation}
which can also be written as
\begin{equation}
    \partial_x Q(x,t) = - \partial_t [\pi R^2(x,t)], \label{eq:SM_h_Q}
\end{equation}
As such, a local drop in $Q(x,t)$ must come accompanied by a local area increase (and viceversa) to enforce incompressibility.

While a deeper analysis of the equations is possible, Eqs. \eqref{eq:SM_h_Q} and \eqref{eq:SM_displacement_final} will be sufficient for the purposes of this manuscript. Because the elastic units will be coarse-grained into single points which store information regarding the volume and the spatially-averaged pressure, there is no need to know the flow field.

\subsection{Active unit}\label{sec:SI_active}

We will now concern ourselves with the active units. As all active units are interchangeable, we pick one of them and shift the frame of reference so that $x=0$ corresponds to the leftmost point of the unit. We again restrict our analysis to thin units 
\begin{equation}
    \frac{R_a}{L_a} \ll 1, \label{eq:SM_thinchannel}
\end{equation}
where $R_a$ and $L_a$ are the spatially-averaged radius of the active units, and their length, respectively
\begin{equation}
    R_a = \frac{1}{L_a} \int\limits_0^{L_a} R(x) dx.
\end{equation} 
We first note that incompressibility of the fluid still holds, and since the walls are rigid ($\partial_t R(x,t) =0)$, from Eq. \eqref{eq:SM_h_Q}, we obtain
\begin{equation}
    \partial_x Q(t) = 0,
\end{equation}
inside the active unit. Next, we shall derive the distribution of solute. Because of the length scale separation between tube radius and height, the solute profile in the radial direction relaxes much quicker than in the longitudinal direction. Let us now determine what this radial profile should be. We restrict our analysis to tubes that are thin enough such that the dominant transport mechanism in the radial direction is diffusion, meaning
\begin{equation}
    Pe_r = \frac{\tau_D}{\tau_A} = \frac{\overline{v}_r R_a}{D}\ll 1,
\end{equation}
where $Pe_r$ is the P\'eclet number in the radial direction, $\overline{v}_r$ is the typical magnitude of the velocity in the radial direction, and $\tau_D = R_a^2/D$ and $\tau_A= R_a/\overline{v}_y $ are the diffusive and advective timescales in the radial direction as well. We further restrict the model to a reaction-limited dynamics, meaning that the Damk\"ohler number $Da_r$ is small
\begin{equation}
    Da_r = \frac{\tau_D}{\tau_{\xi}} = \frac{R_a^2}{D} \left( \frac{\overline{\rho}}{\overline{\xi}} R_a\right)^{-1} \ll 1,
\end{equation}
where $\overline{\xi}$ and $\overline{\rho}$ are the typical magnitudes of $\xi(x)$ and $\rho(x,t)$, and $\overline{\xi}/\overline{\rho}$ is the effective velocity at which the catalyst brings solute into the system, and thus $\frac{\overline{\rho}}{\overline{\xi}} R_a $ is the timescale for "transport" into the channel due to the activity. If the two conditions above are met, we may approximate
\begin{equation}
    \rho(x,r,t) \approx \rho(x,t),
\end{equation}
as diffusion is quick enough to homogenize the effect of advection or of the reaction at the wall. Integrating Eq. \eqref{eq:SM_adv-diff} over the radial direction yields 
\begin{equation}
    \label{eq:SM_packtop}
    \dot{p}(x,t) = - 2\pi \int_{0}^{R(x)} r \left( \frac{1}{r} \partial_r [r j_r(x,r,t)]+ \partial_x j_x(x,r,t) \right) dr  - \chi p(x,t),
\end{equation}
where we define $p(x,t)$ as the integral of $\rho(x,t)$ over the cross-section
\begin{equation}
    p(x,t) = \pi R^2(x) \rho(x,t),
\end{equation}
with units $m^{-3}$. Using Leibniz's rule, we may rewrite Eq. \eqref{eq:SM_packtop} as
\begin{equation}
    \dot{p}(x,t) = - 2 \pi \partial_z \left( \int\limits_0^{R(x)} r j_x(x,r,t) dr \right) + 2\pi R(x)^2 \hat{\xi}(x) - \chi p(x,t), \label{eq:SM_inbetween_p}
\end{equation}
where we have used Eq. \eqref{eq:SM_bc_rho}. Using the expression for the flux (Eq. \eqref{eq:SM_app-J-0}) and the Leibniz rule again, Eq. \eqref{eq:SM_inbetween_p} yields 
\begin{equation}
   \dot{p}(x,t) = -\partial_x \left( \frac{Q(t)}{\pi R^2(x)} p(x,t) \right) - D\partial_x^2 p(x,t) + 2D\partial_x \left( \frac{\partial_x R(x)}{R(x)} p(x,t) \right) + 2\pi R(x)^2 \hat{\xi}(x) - \chi p(x,t),  \label{eq:SM_bigOldEqForp}
\end{equation}
where $Q(t)$ is now independent of $x$. We have obtained a differential equation for the solute concentration which depends on the {current} $Q(t)$. The next step is thus to solve the Stokes equation (Eq. \eqref{eq:SM_stokes}) in order to obtain the velocity profile (and thus $Q(t)$). In the thin-channel regime (Eq. \eqref{eq:SM_thinchannel}), the velocity in the $x$ direction varies much faster in the transversal direction than it does in the longitudinal direction. As such, the Stokes equation (up to terms quadratic in $R_a/L_a$) reduces to 
\begin{equation}
    \eta r^{-1} \partial_r \left[ r \partial_r v_x(x,r,t) \right] =  \partial_z P(x,t),  
\end{equation}
which can be integrated twice in the radial direction to yield
\begin{equation}
    \label{eq:SM_vx_Poi}
    v_x(x,r,t) = v_{a,x}(x,t) - \frac{\partial_z P(x,t)}{4 \eta} \Big[ R^2(x) - r^2 \Big]\,,
\end{equation}
where we have used the diffusioosmotic slip boundary conditions $v_x(x,r=R(x),t) = v_{a,x}(x,t)$, with $v_{a,x}(x,t)$ being the $x$ component of the active slip velocity. Integrating Eq. \eqref{eq:SM_vx_Poi} once again yields
\begin{equation}
    \label{eq:SM_Q_forBC}
    Q(t)=v_{a,x}(x,t) \pi R^2(x)-\frac{\pi}{8}\frac{\partial_x P(x,t)}{ \eta}R^4(x)\,.
\end{equation}
Isolating $\partial_x P(x,t)$ in one side and integrating both sides over the $x$ direction yields
\begin{equation}
Q(t) =  \frac{\pi \Delta P(t)}{8 \eta} \left(\int\limits_0^{L_a} R^{-4}(x) dx \right)^{-1}  + \pi   \left(\int\limits_0^{L_a} R^{-4}(x) dx \right)^{-1} \left(\int\limits_0^{L_a} R^{-2}(x) v_{a,x}(x,t) dx \right), \label{eq:SM_Q_forBC2}
\end{equation}
where 
\begin{equation}
    \Delta P(t) = P(0,t) - P(L_a,t)
\end{equation}
is the pressure drop across the active unit. With this convention, positive pressure drive flows from left to right. We now need an expression for $v_{x,a}(x,t)$. From Eq. \eqref{eq:SM_vslip}, we obtain
\begin{equation}
    v_{a,x} (x,t) = - \frac{\curlyL}{\beta \eta} [ \nabla - \mathbf{n}(x,\phi)  (\mathbf{n}(x,\phi)  \cdot \nabla) ] \rho(x,t) \cdot \mathbf{e}_x,
\end{equation}
where $\mathbf{e}_x$ is a director vector pointing in the positive $x$ direction. Using the definition of the normal vector 
\begin{equation}
    \mathbf{n}(x,\phi) = \frac{1}{\sqrt{1 + (\partial_x R(x))^2}} \left[ - \partial_x R(x) \mathbf{e}_x + \mathbf{e}_r(\phi), \right]
\end{equation}
we obtain
\begin{equation}
        v_{a,x} (x,t) = -\frac{\curlyL}{\beta\eta}\left[ \partial_x\rho(x,t)-\frac{(\partial_x \rho(x,t)) (\partial_x R(x)) + r^{-1}\partial_r(r\partial_r \rho(x,t))}{\sqrt{1 + (\partial_x R(x))^2}} \cdot \frac{\partial_x R(x)}{\sqrt{1 + (\partial_x R(x))^2}}\right].
\end{equation}
which to linear order in $\partial_x R(x)$ yields 
\begin{equation}
        v_{a,x} (x,t) = - \frac{\curlyL}{\beta \eta}  \partial_x \rho(x,t) 
\end{equation}
which when plugged into Eq. \eqref{eq:SM_Q_forBC2} yields 
\begin{equation}
   Q(t) =  \frac{\pi \Delta P(t)}{8 \eta } \left( \int\limits_0^{R(x)} R^{-4}(x) dx\right)^{-1} - \frac{\curlyL}{\beta \eta} \left( \int\limits_0^{R(x)} R^{-4}(x)dx \right)^{-1} \int\limits_0^{L_a} R^{-2}(x) \partial_x[p(x,t) R^{-2}(x)]dx \label{eq:SM_inbetween_Q}
\end{equation}
The first term encodes for the response of a purely passive element ($v_{a,x} = 0$) under an externally applied pressure drop $\Delta P(t)$. The second term encodes for the active flows. To advance analytically, we expand the solute concentration in powers of the corrugation height
\begin{equation}
    p(x,t; \delta R(x)) = p_0(x,t) + \sum\limits_{l >0} p_l(x,t) \delta R(x)^l,
\end{equation}
where 
\begin{equation}
    R(x) = R_a + \delta R(x).
\end{equation}
We focus on active units exhibiting shallow corrugation
\begin{equation}
    \frac{\delta R(x)}{R_a} \ll 1,
\end{equation}
enabling us to safely discard term or order two and above in $\delta R(x) / R_a$. With the above considerations, by making use of the fact that no solute leaks out of the elements ($p(0) = p(L_a) =0$ ) and restricting the shape to obey $R(0) = R(L_a)$, Eq. \eqref{eq:SM_inbetween_Q} yields
\begin{equation}
   Q(t) =  \frac{\pi \Delta P(t)}{8 \eta } \left( \int\limits_0^{R(x)} R^{-4}(x) dx\right)^{-1} + \frac{\curlyL}{\beta \eta} \frac{2R_a}{L_a} \int\limits_0^{L_a} \delta R(x) \partial_x p_0(x,t) dx \label{eq:SM_Qfromdxp}
\end{equation}
where $p_0(t)$ is obtained by plugging-in the expansion in powers of $\delta R/ R_a$ in Eq. \eqref{eq:SM_bigOldEqForp} and keeping to the zeroth-order in $\delta R(x)$ yielding
\begin{equation}
    \dot{p}_0(x,t)  = D \partial_x^2 p_0(x,t) + \frac{Q(t)}{\pi R_a^2} \partial_x p_0(x,t) + 2\pi R(x) \hat{\xi}(x) - \chi p_0(x,t)).
\end{equation}
Assuming instantly fast relaxation of the solute concentration to the underlying flow field $\dot{p}_0(x,t)=0$ and defining
\begin{equation}
   \xi(x) = 2\pi R(x) \hat{\xi}(x)
\end{equation}
\begin{equation}
    \label{eq:SM_forp0}
    0  = D \partial_x^2 p_0(x,t) - \frac{Q(t)}{\pi R_a^2} \partial_x p_0(x,t) + \xi(x) - \chi p_0(x,t).
\end{equation}
As the values $p_l$ for $l>0$ will no longer be considered, we simplify the notation by dropping the subscript and renaming $p_0(x,t)$ as $\mathcal{P}(x,t)$
\begin{equation}
\mathcal{P}(x,t) \equiv p_0(x,t).
\end{equation}
We now perform a Fourier expansion in space of $\mathcal{P}(x,t)$ and $\xi(x)$, such that
\begin{equation}
\mathcal{P}(x,t)=\mathcal{P}_0(t) + \sum\limits_{j>0} \mathcal{P}_j(t) \cos\left( k_j x  \right) +  \sum\limits_{j>0} \tilde{\mathcal{P}}_j(t) \sin\left( k_j x  \right), \label{eq:SM_def-p2}
\end{equation}
\begin{equation}
 \xi(x)=\xi_0 + \sum\limits_{j>0} \xi_j \cos\left( k_j x  \right) + \sum\limits_{j>0} \tilde{\xi}_j \sin\left( k_j x \right)\label{eq:SM_def-xi2},
\end{equation}
where we define the wavelength
\begin{equation}
    k_j = \frac{2  \pi}{L_a}j,
\end{equation}
the Fourier coefficients $\{\mathcal{P}_j\}$ and $\{\mathcal{\xi}_j\}$ associated to the symmetric part of the expansion, and the coefficients $\{\tilde{\mathcal{P}}_j\}$ and $\{\tilde{\mathcal{\xi}}_j \}$ associated to the antisymmetric part of the expansion. Plugging Eqs.~\eqref{eq:SM_def-p2} and~\eqref{eq:SM_def-xi2} in Eq.~\eqref{eq:SM_forp0} leads to
\begin{align}
\dot{\mathcal{P}}_j(t) &= a_j  \mathcal{P}_j(t) + Q(t)b_j\tilde{\mathcal{P}}_j(t) + \xi_j, \label{eq:SM_system_p}\\ 
\dot{\tilde{\mathcal{P}}}_j(t) &= a_j  \tilde{\mathcal{P}}_j(t) - Q(t)b_j \mathcal{P}_j(t)  +\tilde{\xi}_j,\label{eq:SM_system_pTilde}
\end{align}
where
\begin{align}
\label{eq:SM_a_def}
a_j &= - \left[ \chi + D k_j^2 \right], \\
\label{eq:SM_b_def}
b_j &= - \frac{k_j}{\pi R_a^2}.
\end{align}
We insert Eq.~\eqref{eq:SM_def-p2} into Eq.~\eqref{eq:SM_Qfromdxp} to obtain
\begin{equation}
\label{eq:SM_centralQ}
Q(t) =  \frac{\pi}{8 \eta} \frac{\Delta P}{L_a}R_a^4 - \frac{\mathcal{L}}{\beta\eta } \sum\limits_{j=1}^\infty ( \tilde{\mathcal{P}}_j(t) \Gamma_j  - \mathcal{P}_j(t) \Theta_j ),
\end{equation}
where
\begin{align}
\Gamma_j =& - \frac{ k_j}{L_a\mathcal{H}_r}  \left[\int\limits_{0}^{L_a} \delta h (x) \cos\left( k_j x \right) dx \right],\label{eq:SM_defGamma}\\
    \Theta_j = &- \frac{k_j}{L_a\mathcal{H}_r}  \left[\int\limits_{0}^{L_a}  \delta h (x) \sin\left( k_j x \right) dx \right],\label{eq:SM_defTheta}
\end{align} 
and Fourier expanding
\begin{equation}
\delta R(x) = \sum_{j=1}^\infty R_{j}\cos(k_j x) + \sum_{j=1}^\infty \tilde{R}_{j}\sin(k_j x) ,
\end{equation}
we obtain
\begin{align}
\Gamma_j =& - \frac{ k_j}{R_a}  R_j,\label{eq:SM_defGamma2}\\
    \Theta_j = &- \frac{k_j}{R_a}  \tilde{R}_j.
\end{align}
In the case where $Q(t)$ changes gently enough in time, the solute concentration can be considered to adjust adiabatically to the flow rates. In such a case, Eqs.~\eqref{eq:SM_system_p} and ~\eqref{eq:SM_system_pTilde} result in
\begin{eqnarray}
a_j  \mathcal{P}_j(t) + Q(t)b_j\tilde{\mathcal{P}}_j(t)  + \xi_j = 0,\\
a_j  \tilde{\mathcal{P}}_j(t) - Q(t)b_j \mathcal{P}_j(t)  + \tilde{\xi}_j = 0,
\end{eqnarray}
which, if $a_j^2 +b_j^2Q(t)^2 \neq 0$, is solved by 
\begin{eqnarray}
\label{eq:SM_FourierPStart}
\mathcal{P}_j(t) = \frac{Q(t)b_j \tilde{\xi}_j - a_j \xi_j}{a_j^2+b_j^2Q^2(t)},\\
\tilde{\mathcal{P}}_j(t) = -\frac{a_j \tilde{\xi}_j + Q(t) b_j \xi_j}{a_j^2+b_j^2Q^2(t)}.\label{eq:SM_FourierPEnd}
\end{eqnarray}
The solution of Eqs. \eqref{eq:SM_FourierPStart} and \eqref{eq:SM_FourierPEnd} can then be plugged into Eq \eqref{eq:SM_centralQ} to obtain a polynomial equation for the values of $Q(t)$ that are allowed for a given pressure drop $\Delta P(t)$. In Supplementary Material \ref{sec:LBM}, we plot $Q(\Delta P(t))$ for an hourglass-shaped unit with a central patch of catalytic material. Such a setup showcases bistability and hysteresis in the {current}. 

The last thing that must be noted is that we have reduced the active unit to the function $Q(\Delta P)$, meaning that no information is needed from the elastic units (barring the pressure values at the contact points) to fully describe the solute and velocity profile inside the active units. We have seen in the previous section that the time evolution of the elastic units needs information from the active units only at the contact points. We now show how to reduce the fully 3D dynamics discussed in the previous sections to an effective network model solving for the volume and average pressure in each elastic unit.

\subsection{Reduction to network model}\label{sec:SI_lattice}

In this section, we shall reduce the results from the above two sections to a set of dynamical equations in the style of a flow network model \cite{Blunt_2002}. The quantities of interest are the spatially-averaged pressures inside each elastic-walled elements, as well as their volumes. These pressures and volumes change in time due to flow that is imposed by the active elements.
For convenience, we once again change the frame of reference, such that $x = 0$ and $x=L_e$ correspond to the edges of a given elastic unit. We are able to integrate Eq. \eqref{eq:SM_h_Q} in the $x$ direction to obtain
\begin{equation}
    \partial_t V(t) = Q(0,t) - Q(L_e,t), \label{eq:SM_Vdot_fromQ}
\end{equation}
where $V(t)$ is the chamber volume defined as
\begin{equation}
    V(t) = \pi \int\limits_0^{L_e} R^2(x,t) dx.
\end{equation}
In the elastic unit, we identify $R(x,t) = R_1 + u_r(t)$, where $u_r(t)$ is the norm of the displacement vector given in Eq. \eqref{eq:SM_displacement_final}. To linear order in $u_r(t)$ we have
\begin{equation}
    V (t)= \pi R_1^2 L_e + \frac{\pi R_1^3}{2 \delta R} \frac{2\mu_L + \lambda}{(\mu_L + \lambda) \mu_L} L_e [\langle P \rangle(t) - \overline{P}], 
\end{equation}
or
\begin{equation}
     V(t) - V_0 = \gamma (\langle P \rangle (t) -\overline{P}), \label{eq_VfromP_SM}
\end{equation}
where $V_0$ is the volume in the rest configuration ($\langle P \rangle  =\overline{P}$)
\begin{equation}
    V_0 = 2\pi R_1^2 L_e,
\end{equation}
and
\begin{equation}
    \gamma = \frac{\pi R_1^3}{2 \delta R} \frac{2\mu_L + \lambda}{(\mu_L + \lambda) \mu_L} L_e.
\end{equation}
is a proportionality constant between changes in volume and changes in average pressure. Under such a framework, the elastic units are entirely represented by two numbers: their volume and their pressure (which are connected via Eq. \eqref{eq_VfromP_SM}). The volume of an elastic unit evolves in time according to the fluxes imposed on either side from the active units (Eq. \eqref{eq:SM_Vdot_fromQ}). In turn, the pressures of the elastic units (computed from the volumes via Eq. \eqref{eq_VfromP_SM}) determine these fluxes (Fig. \ref{Fig_schematic_ring}). As such, we have a closed system of equations which are solved computationally in Section \ref{sec:discrete}.

\section{Analytical results and scaling arguments used throughout the main text}

\subsection{Overdamped diffusion of  background oscillations}\label{APPENDIX_background_diffusion}

In simulations initialised with random $b_i$ distributions and mechanical equilibrium at each node, $P_i = \bar{P}$, we observe a transitional behavior with oscillations in the pressure profile that evolves towards a flat profile with a solitary wave, see Figs. \ref{Fig_soliton_emergence} and \ref{APENDIX_Fig_soliton_emergence} and the videos within \ref{APPENDIX_videos}. The evolution of these background oscillations can be understood from the equations of the system, recovering the dimensionless equation \eqref{Eq_local_time_evolution_2} and dropping the tildes:
\begin{equation}
\partial_{t}P_{i}(t)=Q_{i-1}\left[\Delta P_{i-1}(t)\right]-Q_{i}\left[\Delta P_{i}(t)\right],
\end{equation}
where the explicit form for the dimensionless current can be written as, 
\begin{equation}
    Q_{i}[\Delta P_{i}(t)] = [ 1+ \tilde{a}\Delta P_{i}(t)^{2}] b_i(t) + \tilde{m}\Delta P_{i}(t), 
\end{equation}
where $\Delta P_{i}(t) = P_i(t) - P_{i+1}(t)$ and $Q_i$ corresponds to the {current} carried by the active unit bridging elastic units $i$ and $i+1$. Taking into account that $b_i(t)=\pm1$ and considering that the quadratic term of the fit is small, we can approximate the current by neglecting the non-linear correction, yielding to the linearized expression
\begin{equation}
Q_{i}\left[(\Delta P_{i}(t)\right] \approx \pm 1 + \tilde{m} \Delta P_{i}(t).
\end{equation}
Outside the domain boundaries (the kink and anti-kink), the constant term $\pm 1$ cancels out between neighboring currents when inserted into the evolution equation. Substituting this approximation into the equation for $\partial_t P_i$, we obtain
\begin{equation}
\partial_{t}P_{i}(t) \approx \tilde{m} \left( \Delta P_{i-1}(t) - \Delta P_i(t) \right) = \tilde{m} \left( (P_{i-1}(t) - P_i(t)) - (P_i(t) - P_{i+1}(t)) \right).
\end{equation}
This expression corresponds to a discrete Laplacian for the pressure field
\begin{equation}
\partial_t P_i(t) \approx \tilde{m} \left( P_{i-1}(t) - 2P_i(t) + P_{i+1}(t) \right).
\end{equation}
Thus, in the linear regime, the pressure field evolves according to a discrete diffusion equation. To analyse the relaxation of the oscillatory modes, we can decompose the pressure field into spatial Fourier modes:
\begin{equation}
P_i(t) = \sum_k \hat{P}_k(t) e^{\mathrm{i} k i},
\end{equation}
where $k$ is the wave number. Inserting this decomposition into the discrete Laplacian yields
\begin{equation}
\partial_t \hat{P}_k(t) \approx -2 \tilde{m} (1 - \cos k) \hat{P}_k(t).
\end{equation}
This shows that each Fourier mode decays exponentially with a rate proportional to $1 - \cos k$. High-frequency modes (large $k$) decay quickly, while low-frequency modes (small $k$) decay much more slowly, as observed in our simulations.

\subsection{Analytical estimation of the speed of an ASW}
\label{APPENDIX_SECTION_VELOCITY}

For the local and non-local cases, we find stable ASWs that move at constant velocities. The velocity is controlled by the flipping dynamics of the active pores at the boundaries of the solitary wave (kinks/anti-kink). If we neglect the small curvature observed in Fig.~\ref{Fig_schematic_ring} \textbf{c} and use a linear relation between {current} and pressure drop where Eq.~\ref{eq:currentDeltaP} simplifies to $Q\left[\Delta P(t)\right]=m\Delta P(t) +b_i(t) \overline{Q}$, the only two free parameters that can affect the velocity are $\omega$ and $m$ (no other parameter appears in the dimensionless equation \eqref{Eq_time_evolution_NORM}).

Limiting ourselves to the local case ($\omega=0$), so as to estimate the velocity of the ASW in the discrete model, we will try to compute the time required for an active pore at the ASW boundary to switch branches.  Thus, we will use the dimensionless equation
\begin{equation}
Q_{i}\left[\Delta P_i(t)\right]=b_i(t)+\tilde{m}\Delta P_i(t),
\label{APPENDIX_Eq_Current_hysteresis_linear}
\end{equation}
where we omit the tildes for the dimensionless variables, and $\tilde{m}=\frac{m \Delta P_c}{\overline{Q}}$, $\Delta P_i(t) = P_i(t) - P_{i+1}(t)$, and $b_i(t)$ denotes the branch of the active pore, $b_i(t)=1$ for the positive branch and $b_i(t)=-1$ for the negative branch.

The velocity of the ASW should be inverse to the time required by the ASW to move one site in the network. To compute this time we consider an ASW formed by a volume accumulation propagating clockwise along the network (see Fig. \ref{Fig_movimiento_soliton} \textbf{c}). Without loss of generality, in the rest of this section we redefine the dimensionless pressure as $P_i = \frac{P_i-\bar{P}}{\Delta P_c}$. We focus on the front and assume, for simplicity, that at $t=0$ we have the following configuration:  $P_{i-1}(0) = 1$, $P_{i}(0) = 0$. We also consider $b_{i-1}(0)=1$ and $b_{i}(0)=-1$, hence, $Q_{i-1}(\left[\Delta P_{i-1}(0)\right]) = 1 + \tilde{m}$ and $Q_{i}(\left[\Delta P_{i}(0)\right]) = -1$. Consequently, our aim is to determine the time needed for $P_{i}$ to increase from $0$ to $1$. In the case of $\omega=0$, Eq.~\ref{Eq_time_evolution_NORM} yields to $\partial_t P_{i}(t)= Q_{i-1}\left[\Delta P_{i-1}(t)\right] - Q_{i}\left[\Delta P_{i}(t)\right]$, and using Eq.~\ref{APPENDIX_Eq_Current_hysteresis_linear} we obtain
\begin{equation}
\partial_t P_i(t) = \left(1 + \tilde{m}(P_{i-1}(t) - P_i(t))\right) - \left(-1 + \tilde{m}(P_i(t) - P_{i+1}(t))\right).
\end{equation}
Assuming that $P_{i-1}(0) = 1$ and $P_{i+1}(0) = 0$, the time evolution for the pressure field is given by 
\begin{equation}
\partial_tP_i\ (t) + 2 \tilde{m} P_i(t) = 2 + \tilde{m}.
\end{equation}
Solving the differential equation, between $t=0$ and $t=\tau$ knowing that $P_i(0)=0$, leads to
\begin{equation}
P_i(\tau) = \frac{2 + \tilde{m}}{2 \tilde{m}} \left(1 - e^{-2 \tilde{m} \tau}\right).
\end{equation}
Thus, the time that takes elastic unit $i$ to go from an initial pressure $P_i(t)=0$ to a pressure $P_i(\tau) =1$ is
\begin{equation}
\tau = \frac{1}{2 \tilde{m}} \ln \left(\frac{2 + \tilde{m}}{2 - \tilde{m}}\right).
\end{equation}
And, since the velocity should be inversely proportional to this time, it should take the form
\begin{equation}
\tilde{c} = \frac{1}\tau{} = \frac{2 \tilde{m}}{\ln \left(\frac{2 + \tilde{m}}{2 - \tilde{m}}\right)},
\end{equation}
or with dimensions,
\begin{equation}\label{APPENDIX_Eq_velocity_theo}
 c = \frac{\Delta x \overline{Q}}{\gamma \Delta P_c} \frac{2 \tilde{m}}{\ln \left(\frac{2 + \tilde{m}}{2 - \tilde{m}}\right)}.
\end{equation}

\begin{figure}
\centering
\includegraphics[width=0.85\linewidth]{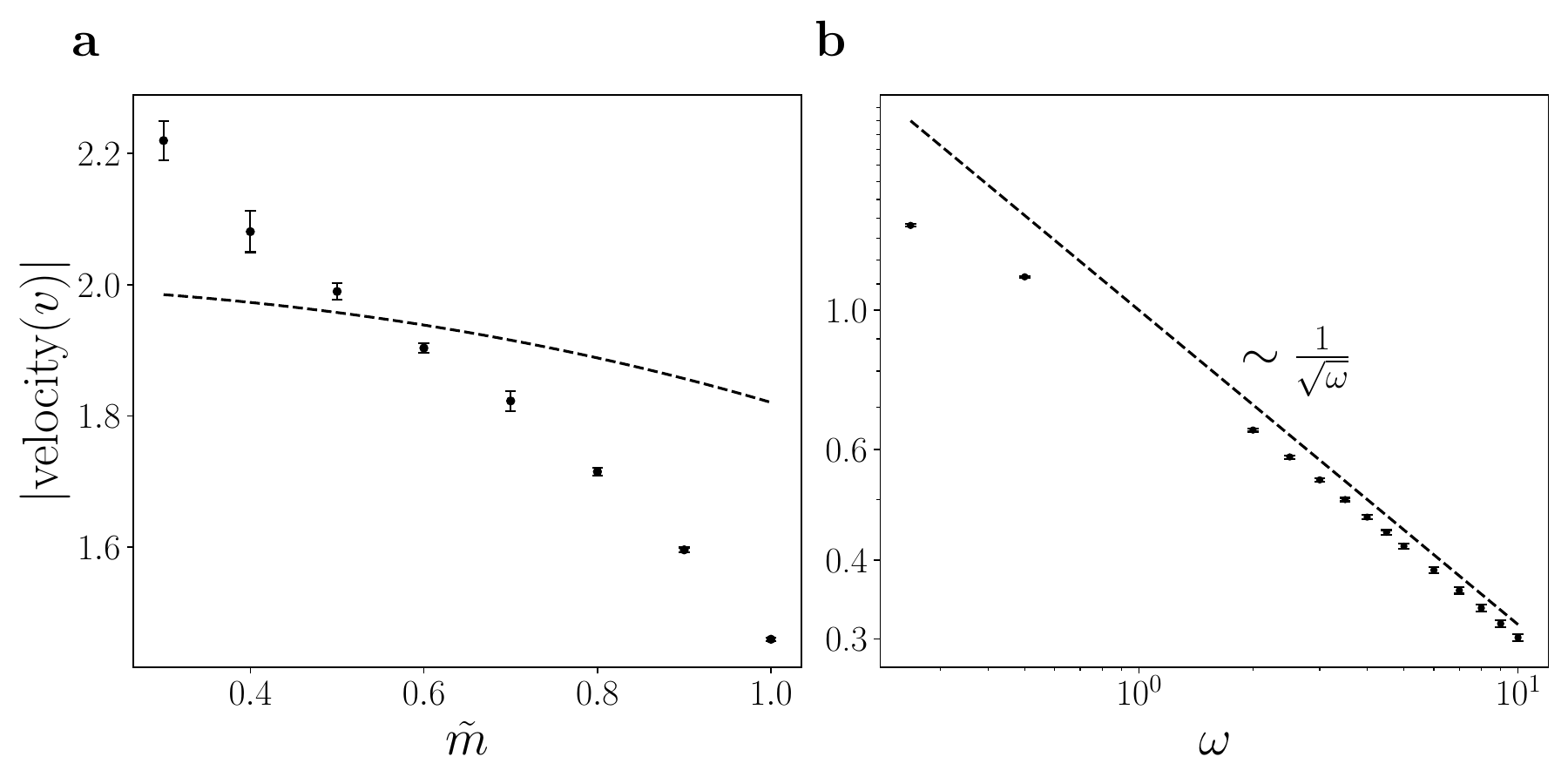}
\caption{
Panel a shows how the velocity of the ASW depends on $m$ at $\omega = 0$, represented by black dots, along with a dashed line indicating an analytical approximation. Panel b displays the velocity dependence on $\omega$, using the default value $m \approx 0.58$ (using the dimensionless expression) obtained from the fit shown in Fig.~\ref{Fig_schematic_ring}-c. Black dots represent the data, and the dashed line illustrates the corresponding scaling behaviour.}
\label{APPENDIX_Fig_velocity}
\end{figure}

Thus, Eq.~\ref{APPENDIX_Eq_velocity_theo} is an approximation for the speed of the ASW that only depends on the slope of the {current} of the active pores. 
Fig.~\ref{APPENDIX_Fig_velocity} \textbf{a} presents the velocity obtained from different simulations (orange dots) with $\omega=0$, together with the analytical estimation, detailed above, as a dashed line. A qualitative agreement is observed, the analytical expression correctly predicts both the magnitude and the decreasing trend, although there is no quantitative match as the velocity obtained from simulations exhibits an almost linear decay.

The speed of the ASW also depends on $\omega$ in the non-local case $(\omega\neq0)$. As shown in Fig.~\ref{APPENDIX_Fig_velocity} \textbf{b}, the velocity of the ASW (orange dots) decreases with increasing $\omega$, with a power law behaviour of $|v| \sim \omega^{-1/2}$. The main reason behind this behaviour is that, as we have discussed in the main text, the height of the ASW scales as $h \sim \omega^{1/2}$, leading to a volume accumulation within the ASW that also scales as $\omega^{1/2}$. Since the currents in the boundaries are independent of $\omega$, the time that it takes to flip the active pores within the boundary to move the ASW one step scales inversely proportional to the height, leading to a velocity that scales as $|v| \sim \omega^{-1/2}$.

\subsection{Introducing the continuum limit: direction of movement and dimensionality analysis of the speed}

Combining non-local volume-pressure coupling, Eq.~\ref{Eq_non_Local_coupling}, and volume conservation Eq.~\ref{Eq_mass_conservation_law}, the time evolution at each elastic unit can be written as,
\begin{equation}\label{APPENDIX_Eq_time_evolution}
\partial_t V_i(t) = \partial_t(\gamma P_i(t)+\alpha\sum_j L_{ij} P_j(t)) = Q_{i-1}\left[\Delta P_{i-1}(t)\right] - Q_{i} \left[\Delta P_i(t)\right].
\end{equation} 
Here $\alpha = \omega\gamma$. We give here a simple argument to understand qualitatively the speed and direction of movement of the ASW using a continuum limit.  Eq.~\ref{APPENDIX_Eq_time_evolution} can be rewritten in the continuous limit identifying each term with its corresponding differential analogue, such that $L_{ij} \leftrightarrow \Delta x^2 \partial_x^2$, and the difference of the currents between two adjacent points in the continuum is given by $\Delta x \, \partial_x Q$, leading to
\begin{equation}\label{APPENDIX_Eq_continuum_limit}
\gamma \partial_tP(x,t)- \alpha \Delta x^2 \partial_t(\partial_x^2 P(x,t))= -\Delta x \partial_x Q(x,t).
\end{equation}
As in the main text, we can render this equation dimensionless using the characteristic scales of the system: $\left[P\right]=\Delta P_{c},\text{ }\left[t\right]=\gamma\Delta P_{c}/\overline{Q},\text{ } \left[x\right]=\Delta x, \text{ }\left[Q\right]=\overline{Q},\text{ } \left[V\right]=\gamma \Delta P_c$ and $\omega=\alpha/\gamma$. Thus, rewriting Eq.~\ref{APPENDIX_Eq_continuum_limit} in terms of these magnitudes we get:
\begin{equation}
    \partial_{\tilde{t}} \tilde{P}(\tilde{x},\tilde{t}) - \omega \partial_{\tilde{x}}^2 (\partial_{\tilde{t}}  \tilde{P}(\tilde{x},\tilde{t}))  = - \partial_{\tilde{x}} \tilde{Q}(\tilde{x},\tilde{t}) 
    \label{APPENDIX_Eq_continuum_limit_normalised}
\end{equation}
In the limit where the volume–pressure coupling is local, $\omega = 0$, Eq.~\ref{APPENDIX_Eq_continuum_limit_normalised} reduces to $\partial_{\tilde{t}} \tilde{P}(\tilde{x},\tilde{t}) = - \partial_{\tilde{x}} \tilde{Q}(\tilde{x},\tilde{t})$. The ASW studied in the main text shows a pressure profile that exhibits a bulge or depression in the same region where the current undergoes a domain transition (see Fig.~\ref{Fig_movimiento_soliton} in the main text). Under this condition, a strong approximation can be made by assuming that the spatial variation of the current is proportional to the pressure gradient, $\partial_{\tilde{x}} \tilde{Q}(\tilde{x},\tilde{t}) \propto \pm \partial_{\tilde{x}} \tilde{P}(\tilde{x},\tilde{t})$ (the $\pm$ sign takes into account the different combinations of pressure and {current} domains), which allows a direct proportional relationship between the time evolution of $\tilde{P}(\tilde{x},\tilde{t})$ and its spatial gradient (Eq.~\ref{APPENDIX_Eq_dtP_propto_dxP})
\begin{equation}\label{APPENDIX_Eq_dtP_propto_dxP}
\partial_{\tilde{t}} \tilde{P}(\tilde{x},\tilde{t})  \propto  \mp \partial_{\tilde{x}} \tilde{P}(\tilde{x},\tilde{t}).
\end{equation}

This equation suggests a constant ASW propagation speed, and using the change of variables $\xi = {\tilde{x}} - \tilde{c} {\tilde{t}}$ and, hence, $\partial_{\tilde{x}} = \partial_\xi,\partial_{\tilde{t}} = -\tilde{c} \partial_\xi$, Eq.~\ref{APPENDIX_Eq_dtP_propto_dxP} may be rewritten as $ -\tilde{c} \partial_\xi \tilde{P}(\xi)  \propto  \mp \partial_\xi \tilde{P} (\xi)$. Therefore, this dimensional analysis provides a speed of the ASW $\tilde{c}=\pm 1$, where the sign depends on the combination of volume accumulation (depletion) and the distribution of currents within the ASW, as we see in Fig. \ref{Fig_movimiento_soliton}. If we give dimensions to the velocity again we obtain,
\begin{equation}
c = \frac{\Delta x \overline{Q}}{\gamma \Delta P_c}.
\end{equation}

\subsection{Height of the ASW}

\label{APPENDIX_height_of_soliton}

Considering only local pressure-volume coupling ($\omega = 0$), the ASW reaches a height of $\Delta P_c$, which corresponds to the pressure difference required at the boundaries to switch the current branch at the edges. This ensures that both the current and pressure domains propagate together. In contrast, when non-local volume-pressure coupling is included ($\omega \ne 0$), the pressure profile becomes smoother at the boundaries, and the solitary wave increases in height. We give here a simple scaling argument to understand the power law behavior of the height as a function of $\omega$.

We recover the dimensionless Eq.~\ref{APPENDIX_Eq_continuum_limit_normalised} and follow the same ansatz, where we assume that once the system is in the ASW solution, we can approximate $\partial_{\tilde{x}} \tilde{Q}(\tilde{x},\tilde{t}) \propto \pm \partial_{\tilde{x}} \tilde{P}(\tilde{x},\tilde{t})$. Applying the change of variables $\xi = {\tilde{x}} - \tilde{c} {\tilde{t}}$  $  \implies\partial_{\tilde{x}} = \partial_\xi\text{,  } \partial_{\tilde{t}} = -\tilde{c} \partial_\xi$,  Eq.~\ref{APPENDIX_Eq_continuum_limit_normalised} can be rewritten as 

\begin{equation}\label{APPENDIX_Eq_pressure_in_xi}
-\tilde{c} \partial_\xi \tilde{P}(\xi) + \omega \tilde{c} \partial_\xi^3 \tilde{P}(\xi) \propto -\partial_\xi \tilde{P}(\xi)
\end{equation}
Thus, integrating Eq.~\ref{APPENDIX_Eq_pressure_in_xi} once with respect to $\xi$ yields,
\begin{equation}\label{APPENDIX_Eq_curvature_propto_w}
    \partial_\xi^2 \tilde{P}(\xi) \propto \frac{1}{\omega},
\end{equation}
which reveals that, for $\omega \ne 0$, the ASW profiles become smoother, exhibiting a curvature that scales as $\propto \frac{1}{\omega}$. On the other hand, although the pressure profile becomes smoother at the boundaries, the pressure difference between adjacent reservoirs within the solitary wave boundary must still reach the critical value ($\Delta P/\Delta P_c=1 $, in the dimensionless case), which is the pressure drop required to reverse the direction of the current.

Given that the curvature scales as $1/\omega$ (see Eq.~\ref{APPENDIX_Eq_curvature_propto_w}), and that the pressure difference must reach the critical value $1$ at its local maximum (the center of the boundary), it follows that $\partial_\xi \tilde{P}(\xi)$ should be a function whose maximum value is constant (independent of $\omega$) and whose curvature scales as $1/\omega$. The simplest ansatz satisfying these conditions is
\begin{equation}\label{APPENDIX_Eq_anszat}
    \partial_\xi \tilde{P}(\xi) \propto  e^{-\frac{\xi^2}{\omega}}
\end{equation}
which allows us to compute the height ($h$) of the ASW as 
\begin{equation}
h(\omega) \propto \int_{-\infty}^{\infty} \partial_\xi \tilde{P}(\xi) \, d\xi = \int_{-\infty}^{\infty}  e^{-\frac{\xi^2}{\omega}} \, d\xi \propto \sqrt{\omega},
\end{equation}
which implies that $h(\omega)$ scales as $\sqrt{\omega}$, as we see in the simulations (see Fig.~\ref{Fig_vida_soliton} \textbf{a} of the main text).

\subsection{Lifetime of the Solitary Waves}\label{APPENDIX_sec_SM_Lifetime}

Numerical simulations show that, in the $\omega = 0$ limit, the ASW persist throughout the entire simulation time without changing its shape. In contrast, for $\omega \ne 0$, their lifetime becomes finite and is controlled by the value of $\omega$ and their initial width. As time evolves, ASW boundaries seem to attract each other. To understand the origin of this effective attraction between both boundaries, we turn to an energy-based analysis.
We use Eq.~\ref{Eq_non_Local_coupling} in dimensionless form (without including tildes to avoid cluttering the equation). Without loss of generality, in the rest of this section we redefine the dimensionless pressure as $P_i = \frac{P_i-\bar{P}}{\Delta P_c}$, and assume $\overline{V} = \gamma \Delta P_c$ for simplicity, we get 
\begin{equation}
  P_i(t)  = \sum_j\left[  \delta_{ij} + \omega L_{ij} \right]^{-1} (V_i(t) - 1),
\end{equation}
which is equivalent to a system of springs acting as pistons at each elastic unit (where pressure is controlled by the force on the spring and volume by its strain). In the case $\omega = 0$, the springs are uncoupled, and the pressure at each reservoir is directly proportional to the volume increment. In contrast, when $\omega \ne 0$, the system exhibits non-local coupling between the springs, meaning that the pressure at a given reservoir depends not only on its own volume change but also on the state of its neighbors. 

Since the classical harmonic spring force is given by $F = -k x$, derived from the gradient of an elastic potential energy of the form $E = \frac{1}{2} k x^2$, an analogous interpretation can be applied to the system discussed in this manuscript. Specifically, at any time $t \equiv t_0$, the energy associated with the deformation of the elastic units can be expressed as a quadratic form:
\begin{equation}
    E = \frac{1}{2} \sum_{ij} (V_i - 1) \left[  \mathbf{1} + \omega \mathbf{L} \right]^{-1}_{ij} (V_j - 1) =  \frac{1}{2} \sum_{ij} P_i \left[  \mathbf{1} + \omega \mathbf{L} \right]_{ij} P_j
\end{equation}
which may be decomposed in two contributions, 
\begin{equation}
\label{eq_ener_omega}
E = \frac{1}{2} \sum_{i} P_i^2 + \frac{1}{2} \sum_{ij} P_i \left[  \omega \mathbf{L} \right]_{ij} P_j = E_0 + E_{\omega}
\end{equation}
where, $E_0$ corresponds to the local (uncoupled) elastic energy, while $E_{\omega}$ accounts for the non-local coupling introduced by the parameter $\omega$.
\begin{figure*}
    \centering
    \includegraphics[width=\linewidth]{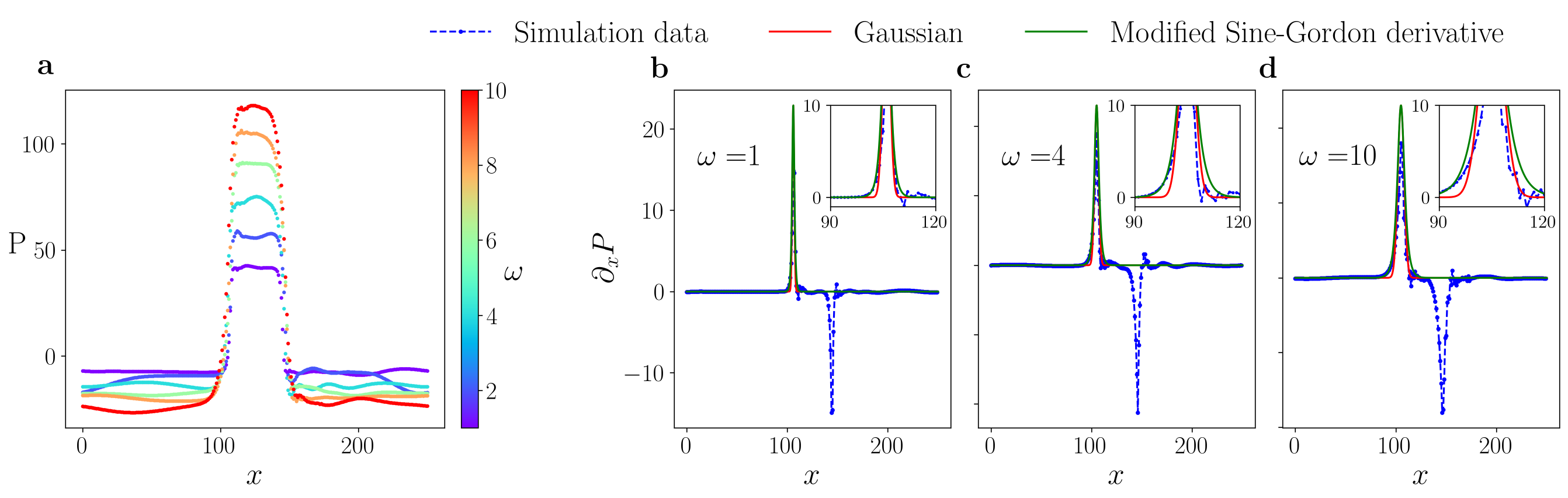}    
    \caption{Panel a displays the pressure profiles of different ASW for several values of $\omega$ (from $\omega = 1$ in dark blue to $\omega = 10$ in red). Panels b, c, and d show the spatial derivative of the pressure ($\partial_x P(x)$) for $\omega = 1$, 4, and 10, respectively. The derivative obtained from simulations is shown in blue, while the red and green curves represent fits to a Gaussian and a modified Sine-Gordon profile, respectively.}
    \label{fig_solitons_diff_omegas}
\end{figure*}
Although, $E_0$ increases as the ASW widens, we observe that the ASW can propagate without changing its shape when $\omega = 0$. Therefore, we assume this term does not create any effective interaction between the boundaries. When $\omega \neq 0$, an additional term, $E_{\omega}$, arises from the non-local coupling of the network’s elastic units. This energy also increases when the ASW widens and provides a plausible explanation for the finite lifetime observed in our simulations.

In order to estimate how the lifetime depends on this expression, we use an analytical ansatz for the shape of the ASW. As in previous sections, we could use the expression $\partial_x P \sim   e^{-\frac{x^2}{\omega}}$ for the boundaries. This approach actually gives approximately the correct scaling, but to get even better results we decided to directly fit the numerical simulations to another function (a modification of the Sine-Gordon soliton solution) that provides a better fit:
\begin{equation}
\label{eq_modified_SG}
\partial_x P(x)= \frac{4 \gamma_1}{\cosh\left(\gamma_2 (x - x_0)\right)},
\end{equation}
where
\begin{align}
    v &= \sqrt{1 - \frac{16}{\Delta P_c^2}}, \label{eq:v_def}\\
    \gamma_1 &= \frac{1}{\sqrt{1 - v^2}}, \label{eq:gamma1_def}\\
    \gamma_2 &= \frac{1}{\sqrt{\omega}}. \label{eq:gamma2_def}
\end{align}

In Fig. \ref{fig_solitons_diff_omegas} \textbf{a} we represent different ASW for different values of $\omega$, and in panels \textbf{b-d} we represent $\partial_x P$. We observe how the Gaussian (red line) is a good approximation close to the peak but does not capture the tail correctly. Whereas the modified sine-Gordon solution Eq.~\ref{eq_modified_SG} approximates better the tail of the simulated peak (blue lines in panels \textbf{b-d} of Fig. \ref{fig_solitons_diff_omegas}). Note also that the simulation results do not show a completely symmetric peak with respect to its center.

\begin{figure}
\centering
\includegraphics[width=\linewidth]{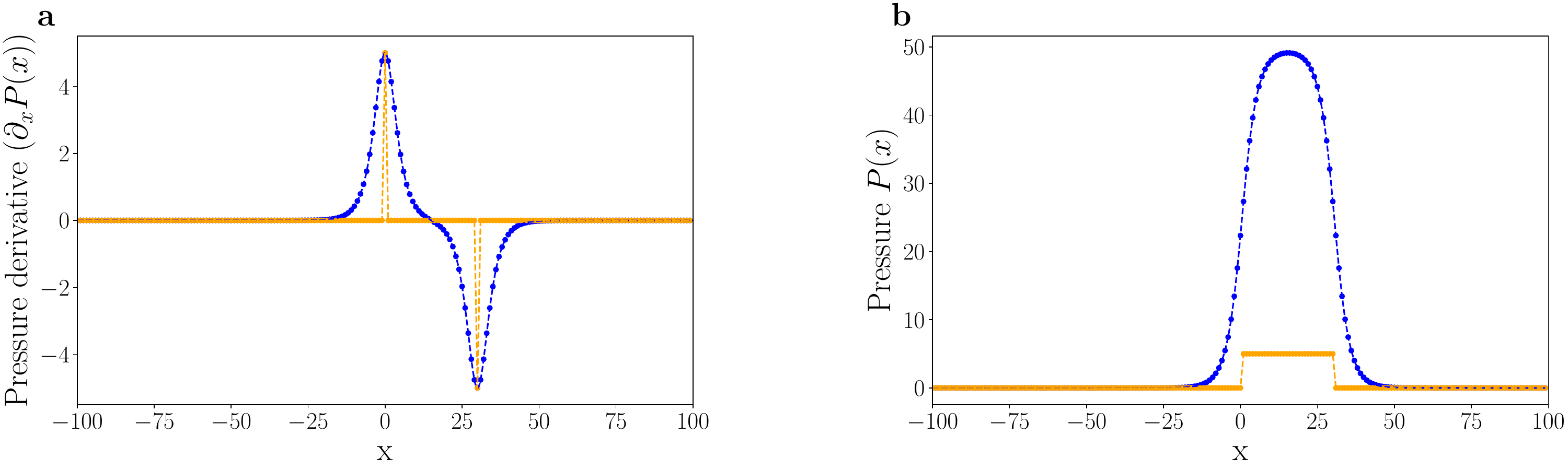}
\caption{Panel a displays the spatial derivative of the pressure, and panel b shows the resulting pressure profile computed from Eq.~\ref{eq_ener_omega}, for $\omega = 10$ (blue) and $\omega = 0$ (yellow) and width $W = 30$.}
\label{APPENDIX_Fig_ASLS_different_widths}
\end{figure}

\begin{figure}
\centering
\includegraphics[width=\linewidth]{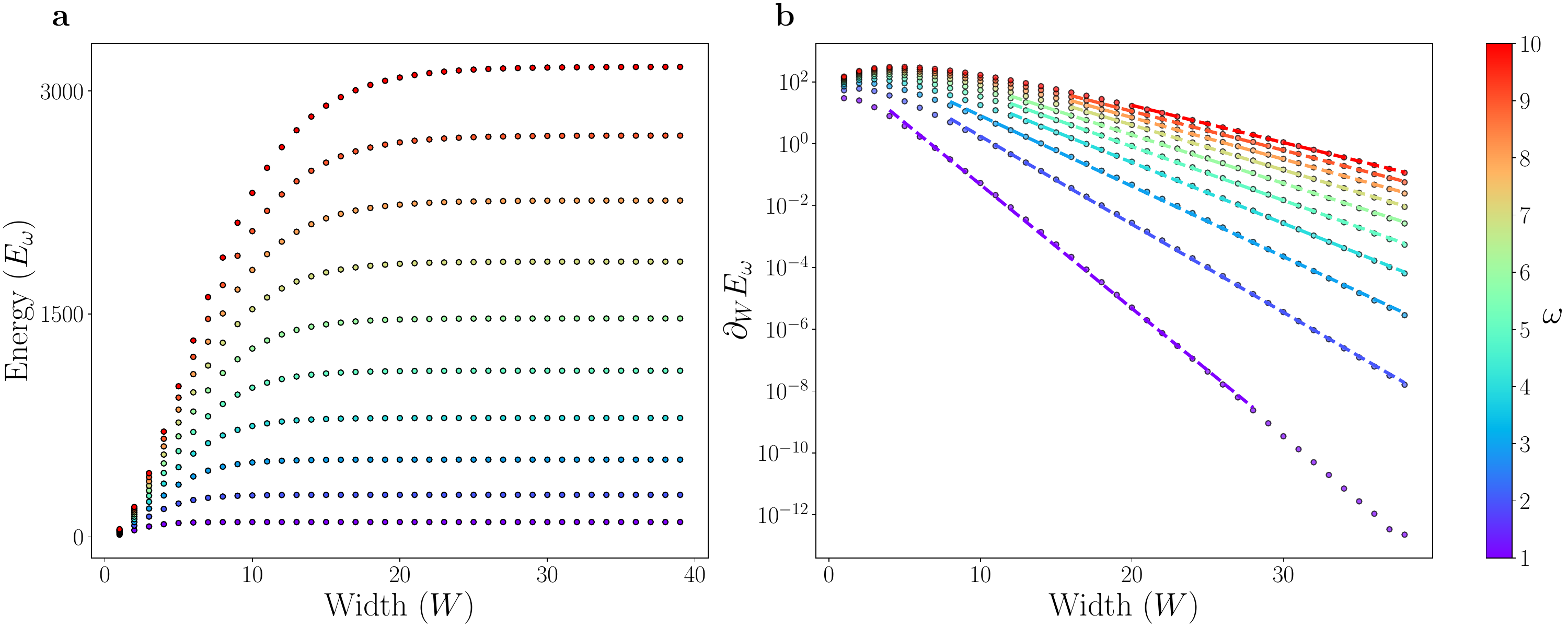}
\caption{Panel a displays the ASW energy ($E_\omega$) as a function of width ($W$) for different values of $\omega$, represented by a colour gradient ranging from blue ($\omega = 1$) to red ($\omega = 10$), which is consistently used in panels b and c. Panel b shows the critical width ($W_c$) as a function of $\omega$, with simulation results indicated by coloured dots and a linear fit on a log-log scale. Panel c presents the derivative of the energy with respect to width for various $\omega$ values, following the same colour scheme as in panel a; it also includes a linear fit of the exponential decay.}
\label{APPENDIX_Fig_energy_SINE_GORDON}
\end{figure}

To compute the energy corresponding to $E_\omega$ for a ASW of different widths and $\omega$, we first use Eq. \eqref{eq_modified_SG} to create two boundaries that lead to a ASW, see Fig. \ref{APPENDIX_Fig_ASLS_different_widths}, where in orange is represented the local case ($\omega=0$) and in blue the case with $\omega\neq 0$. Then we use Eq. \eqref{eq_ener_omega} to compute $E_\omega$ for different values of width $W$ and $\omega$, see Fig.~\ref{APPENDIX_Fig_energy_SINE_GORDON} \textbf{a}. 

This energy, $E_{\omega}$, asymptotes for large values of $W$, suggesting the existence of a decaying attraction between the boundaries of the ASW. Remarkably, Fig.~\ref{APPENDIX_Fig_energy_SINE_GORDON} \textbf{b} illustrates $\partial_W E_\omega$ as a function of $W$, showing how the effective force would decay exponentially with increasing $W$. To test if this effective force would lead to the exponential lifetime observed in the main text, we make another ansatz to integrate the dynamics of the ASW. We assume that the ASW width evolves as an overdamped system subjected to a force derived from the energy $E_\omega$, such as $\partial_t W(t)\propto -\partial_W E_\omega(t)$.

\begin{figure}
\centering
\includegraphics[width=0.5\linewidth]{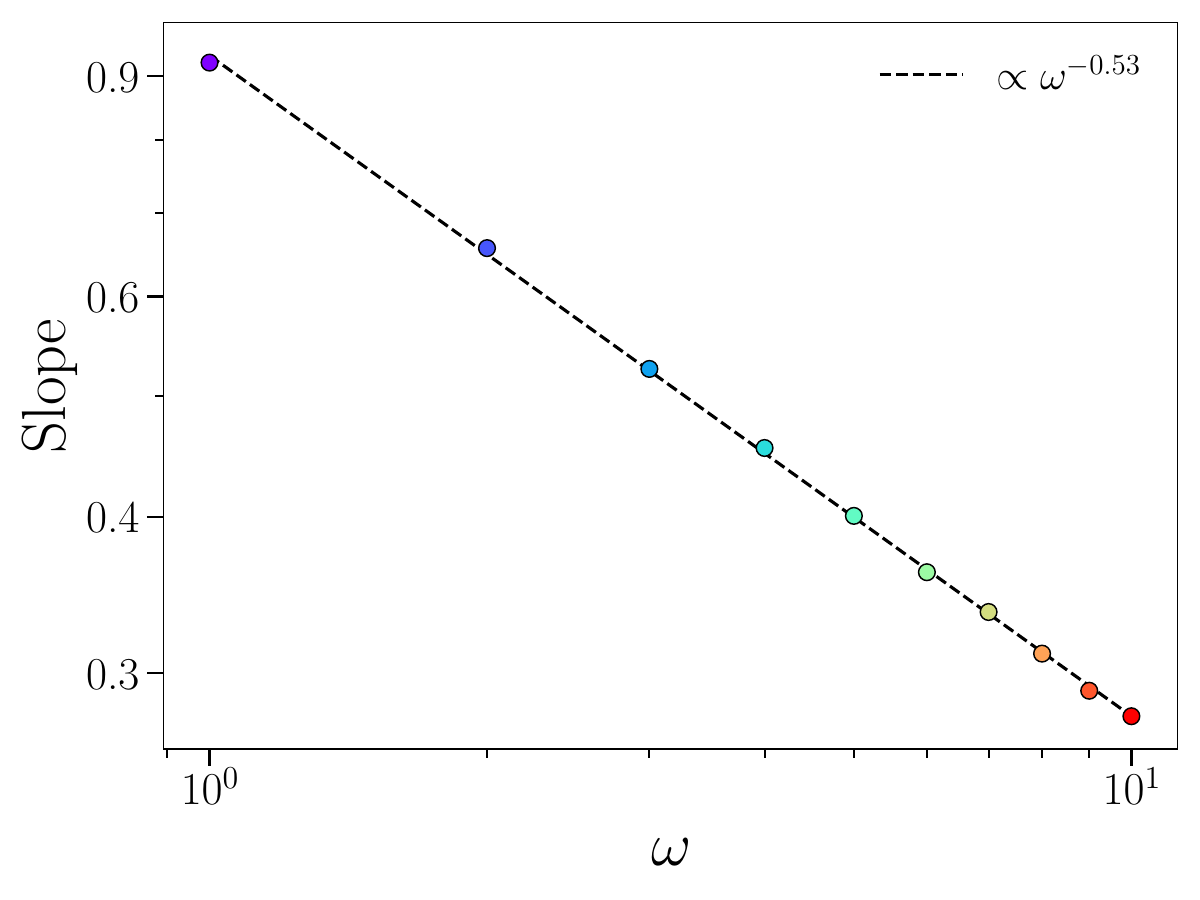}
\caption{Power law behavior of the slopes obtained from the fits shown in Fig.~\ref{APPENDIX_Fig_energy_SINE_GORDON} (coloured dots, using the same colour gradient as in the referenced figure).}
\label{APPENDIX_Fig_slope_exp_fit}
\end{figure}

Since the slopes obtained from the exponential fits behave as $\propto 1/\sqrt{\omega}$, shown in Fig.~\ref{APPENDIX_Fig_slope_exp_fit}. The exponential dependence leads to the relation
\begin{equation}
    \partial_t W(t)\propto -\partial_W E_\omega(t)\propto e^{-W(t)/\sqrt{\omega}} ,
\end{equation}
which can be integrated from $t=0$ to $t=\tau$, yielding
\begin{equation}
    \sqrt{\omega} \left(e^{\frac{W_0}{\sqrt{\omega}}} - e^{\frac{W(\tau)}{\sqrt{\omega}}}\right) \propto \tau.
\end{equation}

The disappearance time of an ASW, denoted $\tau_F$, can be estimated by evaluating the condition $W(\tau_F) = 0$, and assuming $W_0 \gg \sqrt{\omega}$. This leads to the asymptotic expression
\begin{equation}
    \tau_F \propto \sqrt{\omega} \left(e^{\frac{W_0}{\sqrt{\omega}}} - 1\right)
    \sim \sqrt{\omega} e^{\frac{W_0}{\sqrt{\omega}}},
\end{equation}
which is consistent with the behaviour observed in the numerical simulations: in Fig. \ref{Fig_vida_soliton} \textbf{b} the exponent $\beta$ scales as $1/\sqrt{\omega}$.

\section{Additional simulations in the discrete active flow network}

\subsection{Probability of the spontaneous emergence of Active Solitary Waves}

\label{SM_ASW_probability}

\begin{figure}
\centering
\includegraphics[width=0.5\linewidth]{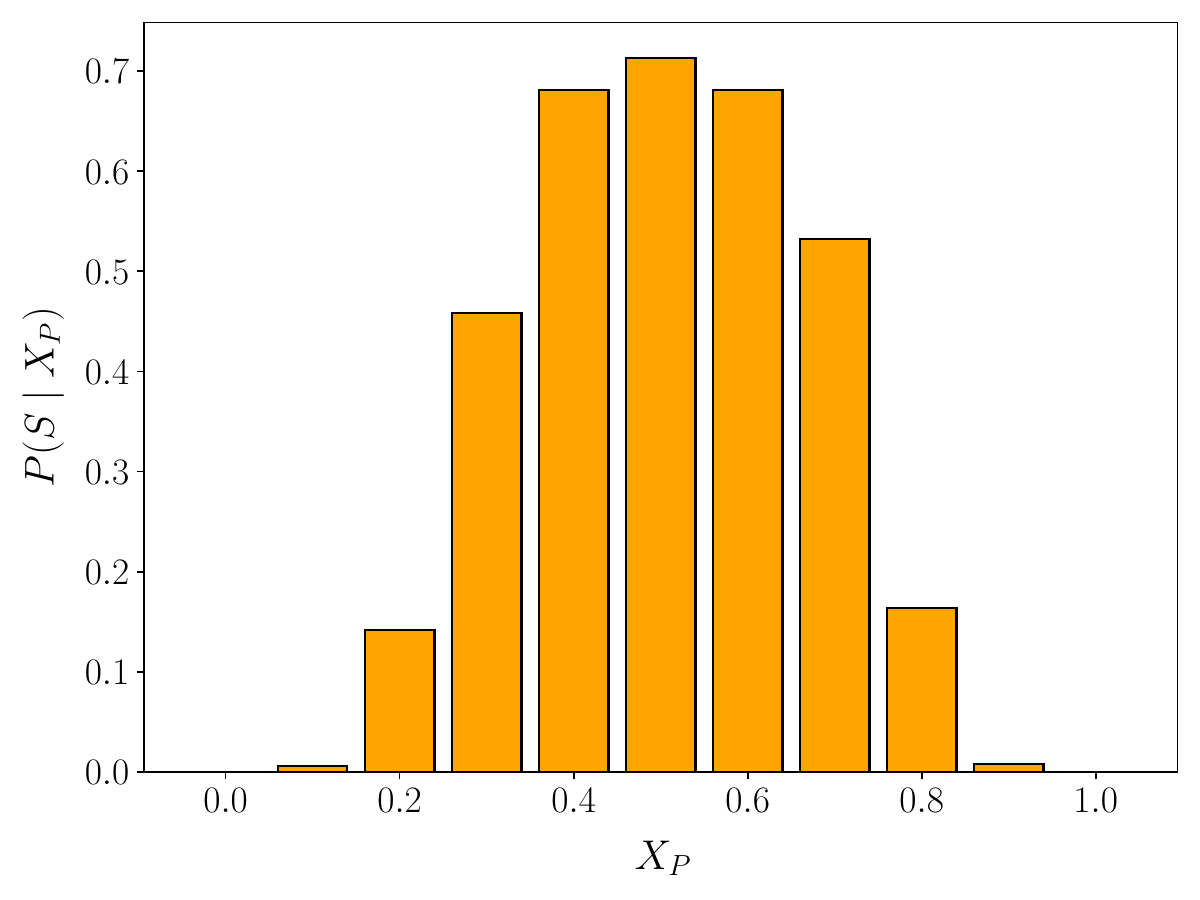}
\caption{ Probability of the 
spontaneous emergence of Active Solitary Waves (ASW), considering only local coupling ($\omega = 0$). The probability $P(S \mid X_P)$ is shown as a function of the initial concentration of active pores pointing clockwise, i.e., along the positive branch ($b_i(t=0) = + 1$), denoted by $X_P$. Simulations are carried out using the dimensionless equations, with relevant parameter values:  $N = 251$, \textcolor{black}{$\overline{Q} \approx 4.61, m \approx 0.117$, $a \approx 2\cdot 10^{-4}$ and $\Delta P_c\approx 23.04$}.
}
\label{APPENDIX_Fig_probability}
\end{figure}

As mentioned in the main text, the ring can be initialized at mechanical equilibrium, $P=\bar{P}$ in each elastic unit, and a random distribution of flow direction in the active units $b_i=\pm 1$. To generate this random initial configuration, in the main text we assigned the values $b_i(t=0) = \pm 1$ with equal probability. We then define the concentration of initially positive branches as $X_P = N_P / N$, where $N_P$ is the number of active pores initially oriented in the positive direction $b_i(t=0) = + 1$, and $N$ is the total number of active pores in the system.

The probability of the emergence of an ASW depends on the initial value of $X_P$ as shown in Fig.~\ref{APPENDIX_Fig_probability}, where $P(S \mid X_P)$ is plotted as a function of $X_P$. The results follow an approximately symmetric distribution, vanishing at the boundaries $X_P = 0$ and $X_P = 1$, which correspond to trivial states where the flow remains stable along the ring in either the counter-clockwise or clockwise direction, respectively. However, for initial values of $X_P \in [0.4, 0.6]$, $P(S \mid X_P)$ exhibits a broad peak, reaching a maximum at $X_P = 0.5$, where the probability is $P(S \mid 0.5) \approx 0.7$. This highlights the robustness of the spontaneous emergence of Active Solitary Waves (ASW) across a wide range of initial conditions.

\subsection{Active Solitary Waves also emerge spontaneously for the case with non-local coupling ($\omega \ne 0$)}

\begin{figure*}
\centering
\includegraphics[width=\linewidth]{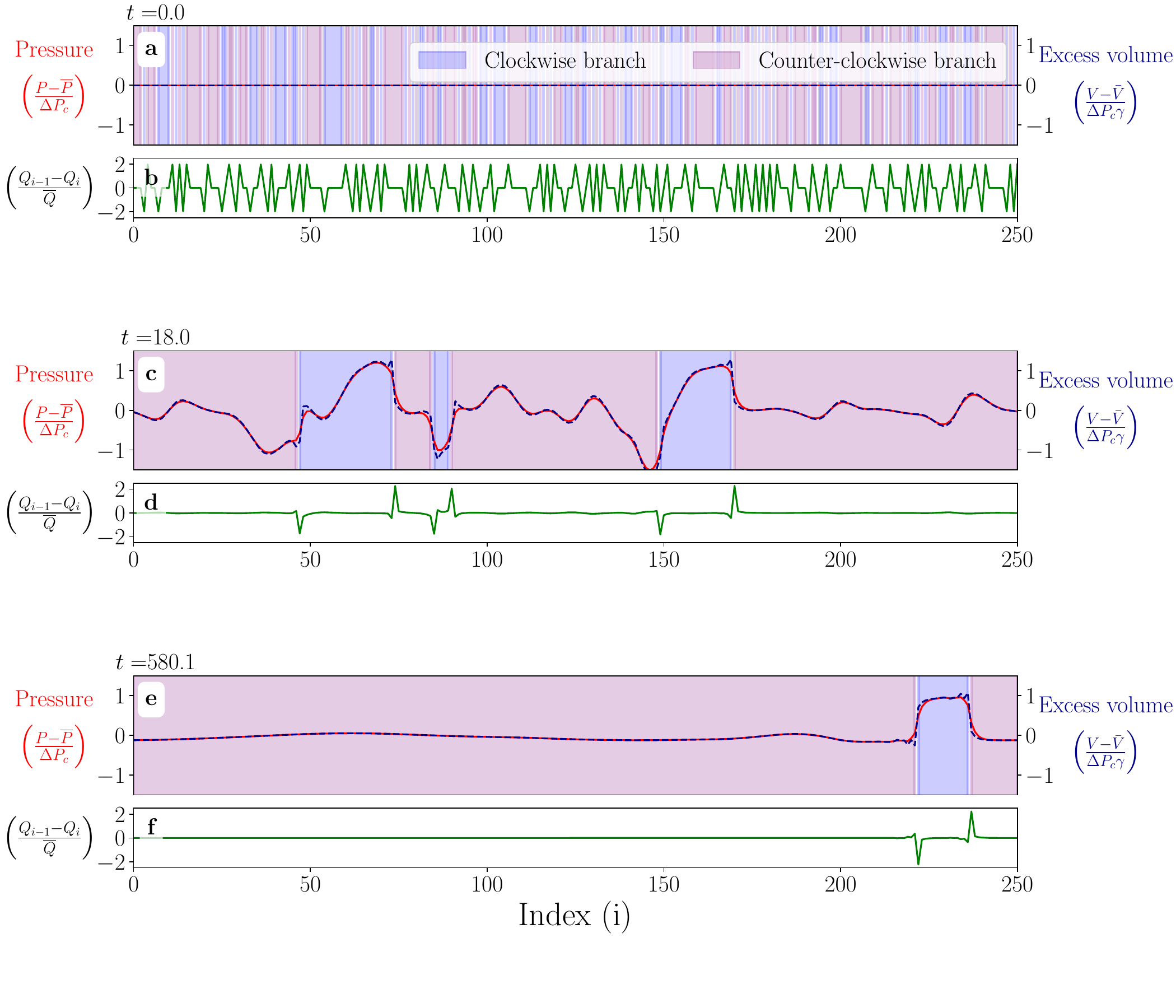}
\caption{Time evolution (from top to bottom) of the system with $b_i$ randomly initialized and producing an ASW, using non-local coupling $\omega=1$. The pressure $P$ (solid red line), volume $V$ (blue dashed line), and net current of each reservoir (green solid line) are depicted. Additionally, the background color represents the branch of the active pores, $b=1$ (clockwise, blue) and $b=-1$ (counter-clockwise, red). The simulation was carried out using the dimensionless equations, with relevant parameter values:  $N = 251$, $\overline{Q} \approx 4.61, m \approx 0.117$, $a \approx 2\cdot 10^{-4}$ and $\Delta P_c\approx 23.04$}
\label{APENDIX_Fig_soliton_emergence}
\end{figure*}

When non-local coupling is included, the Active Solitary Waves still emerge spontaneously as shown in Fig.~\ref{APENDIX_Fig_soliton_emergence}, following the same format as in Fig.~\ref{Fig_soliton_emergence} of the main text.
The main differences from the local case ($\omega = 0$) are: the time needed by the system to relax the oscillations in the pressure profile that appear in the background is shorter; and the final square-shape of the ASW is smoother in the pressure field.

\subsection{Signal transmission in an open system via Active Solitary Waves}

\label{APPENDIX_SM_open_system}

\begin{figure*}
\centering
\includegraphics[width=\linewidth]{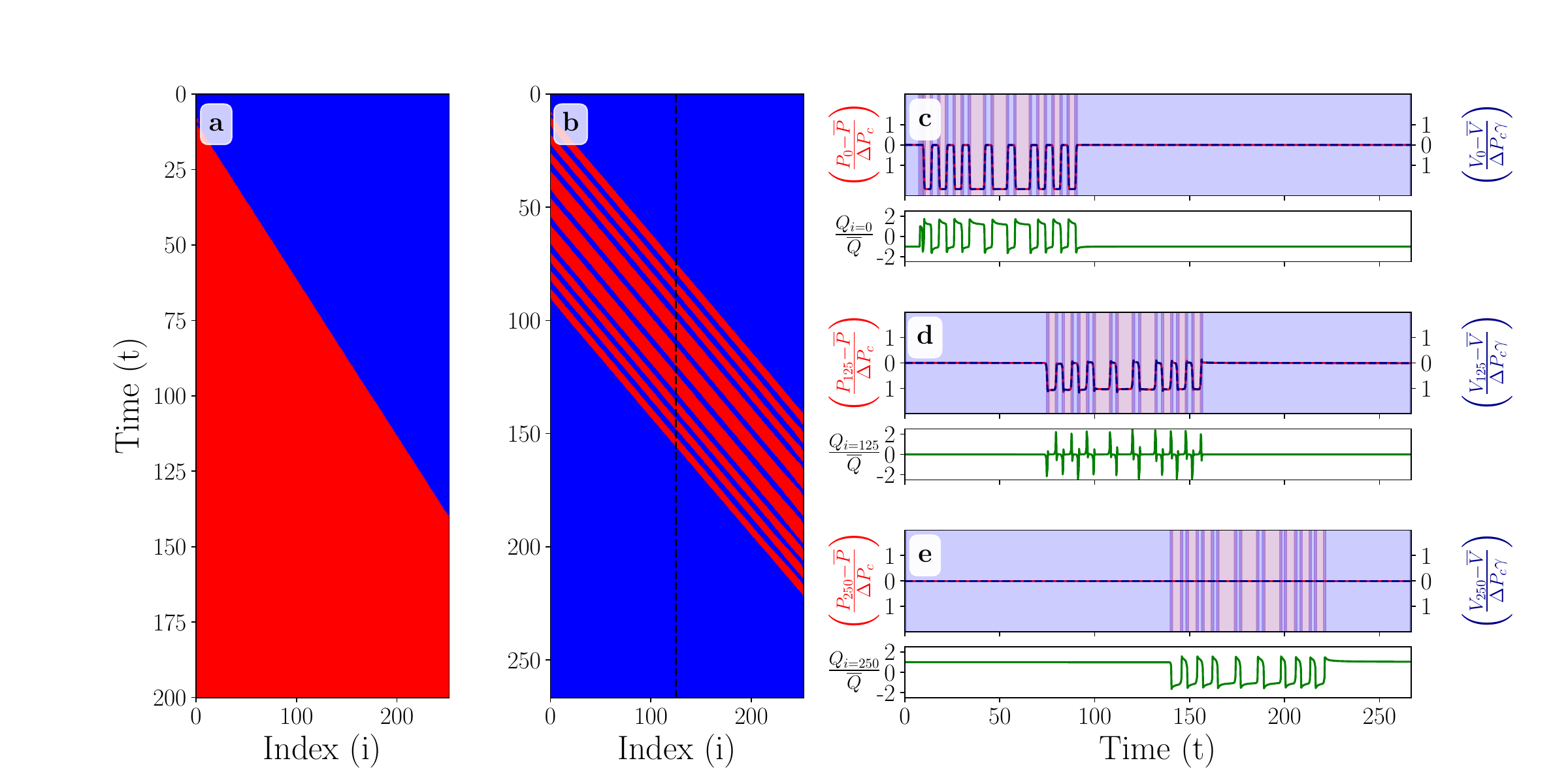}
\caption{Panels a and b: Active pore branch distribution ($b=1$ in blue, $b=-1$ in red) over time (from top to bottom) of a network connected to sources. Panel a shows the system’s response to a single sharp Gaussian signal, which induces a complete polarization reversal. Panel b shows a sequence of signals encoding the message ``SOS'' in Morse code. Moreover, the vertical dashed lines indicates the times at which we show the snapshots presented in panels \textbf{c}, \textbf{d} and \textbf{e}, following the color code of previous figures. These simulations were carried out using the dimensionless equations, with relevant parameter values:  $N = 251$, $\overline{Q} \approx 4.61, m \approx 0.117$, $a \approx 2\cdot 10^{-4}$ and $\Delta P_c\approx 23.04$
}
\label{APPENDIX_Fig_sending_signals}
\end{figure*}

Since ASWs can emerge and propagate in the ring formed by active and elastic units, we would like to consider here an open system. A 1D network connected to two sources, in such a way that a pressure drop may be introduced at one end and generate a ASW that travels against the general flow and is received at the opposite source. In order to test this, we use a 1D network of $N=251$ active and elastic units, with its extremes connected to sources, where we can control the pressure.

Fig.~\ref{APPENDIX_Fig_sending_signals} \textbf{a} shows the evolution of the active units $b_i(t)$ over time (from top to bottom), and consequently the flow domains, when we decrease the pressure at node $i=0$ and keep it at this low value to change the direction of flow in the whole network. Initially, all active units are oriented in the positive direction, corresponding to $b_i(0)=1~\forall i$ (blue in Fig.~\ref{APPENDIX_Fig_sending_signals} \textbf{a}), generating a stable flow in that direction, with the network boundaries, connected to sources, fixed at $P_0(0)=P_{250}(0)=0$. Keeping $P_{250}(t)=0 ~\forall t$, at time $t \approx 10$, a sharp pressure drop is introduced at $i=0$, which causes the active units on the left network's boundary to switch from $b_0(t = 0)=1$ to $b_0(t \approx 10)=-1$ (red in Fig.~\ref{APPENDIX_Fig_sending_signals} \textbf{a}). The pressure drop propagates along the system as a single boundary of a ASW, causing the branch switch of the active units, and after some time ($t\approx140$) it inverts the global polarization of the system, resulting in a steady flow in the negative direction, with all active units in the state $b_i(t>140)=-1~\forall i$.

Thus, an input sequence consisting of two inverted pressure changes, one negative followed by one positive, applied to an initial configuration where $b_{i}(0) = 1~\forall i$, should generate an Active Solitary Wave (ASW) that propagates through the entire network toward the opposite source. This mechanism could enable the transmission of information across the system, for example by encoding messages in Morse code, as illustrated in Fig.~\ref{APPENDIX_Fig_sending_signals} \textbf{b}, where the message ``SOS'' is successfully transmitted via a sequence of pressure pulses (see video at \ref{APPENDIX_videos}). In Fig.~\ref{APPENDIX_Fig_sending_signals} \textbf{c–e}, a sequence of snapshots of the system presented in Fig.~\ref{APPENDIX_Fig_sending_signals} \textbf{b} is shown, following the same format as Fig.~\ref{Fig_soliton_emergence}. Panel \textbf{c} displays the time evolution of elastic unit $0$, which is connected to the source where the pressure is externally controlled to input the desired signals. Panel \textbf{d} corresponds to the time evolution of elastic unit $125$, highlighted with a dashed line in panel \textbf{b}. Note how although we input signals with height $h<-1$ at $i=0$, they quickly evolve to display $h=-1$ as predicted by our theory. Finally, panel \textbf{e} shows the time evolution of elastic unit $250$, which is connected to a source that fixes the pressure to zero at all times. Although no pressure variation occurs at this unit, the net current changes, enabling the signal to be detected at this boundary.

\subsection{Complementary videos} \label{APPENDIX_videos}

We have created several videos to help visualize the different phenomena discussed in the main text:

\begin{itemize}
    \item \textbf{Vid\_emerging\_ASW.mp4}: It illustrates the spontaneous emergence of an ASW at $\omega=0$, following the same colour code and format of one of the panels of figure ~\ref{Fig_soliton_emergence} of the main text.
    \item \textbf{Vid\_emerging\_ASW\_non\_local.mp4}: It illustrates the spontaneous emergence of an ASW at $\omega=1$, following the same color code and format of one of the panels of figure~\ref{Fig_soliton_emergence} of the main text.
    \item \textbf{Vid\_ASW\_death.mp4}: It illustrates the  extinction at $\omega=1$, following the same colour code and format of one of the panels of figure ~\ref{Fig_soliton_emergence} of the main text.
    \item \textbf{Vid\_ASW\_antiASW\_collision.mp4}: It illustrates the annihilation between two ASWs of different sign at $\omega = 1$, following the same colour code and format as one of the panels in Figure~\ref{Fig_soliton_emergence} of the main text.
    \item \textbf{Vid\_multiple\_ASW.mp4}: It illustrates the coexistence of two ASW in the system with $\omega = 0$, following the same colour code and format as one of the panels in Figure~\ref{Fig_soliton_emergence} of the main text.
    \item \textbf{Vid\_sending\_ASW.mp4}: It shows the ASW at $\omega = 0$, generated and propagating through the connected-to-sources network, encoding ``SOS'' in Morse code, using the same format as in the previous videos.
\end{itemize}


\end{document}